\documentclass[journal]{IEEEtran}
\usepackage{amsmath}
\usepackage{amsfonts}
\usepackage{mathtools}
\usepackage{enumitem}
\usepackage{hyperref}
\usepackage[T1]{fontenc}
\usepackage{cite}

\usepackage{siunitx}

\ifCLASSINFOpdf
\else
\fi

\usepackage{graphicx}
\graphicspath{ {./images/} }

\usepackage{boldline,multirow}
\usepackage{mathtools}
\usepackage{algorithm,algorithmic}
\usepackage{subcaption}
\usepackage{comment}

\usepackage{xcolor}

\newcolumntype{R}[1]{>{\raggedleft\arraybackslash}p{#1}}
\newcommand*{\ditto}{\texttt{"}}

\begin{document}
%

\title{A Modular and Transferable Reinforcement Learning Framework for the Fleet Rebalancing Problem}


\author{Erotokritos Skordilis, Yi Hou, Charles Tripp, Matthew Moniot, Peter Graf, David Biagioni
    \thanks{All authors are employed at the National Renewable Energy Laboratory, Golden, CO 80401 USA.  E. Skordilis (Erotokritos.Skordilis@nrel.gov), C. Tripp (Charles.Tripp@nrel.gov), P. Graf (Peter.Graf@nrel.gov), and D. Biagioni (dave.biagioni@nrel.gov) are with NREL's Computational Science Center.  Y. Hou (Yi.Hou@nrel.gov) and M. Moniot (Matthew.Moniot@nrel.gov) are with NREL's Center for Integrated Mobility Sciences.}
    \thanks{The authors would like to thank Devon Sigler and Juliette Ugurimuera for their help in reviewing the manuscript.  This work was authored by the National Renewable Energy Laboratory, operated by Alliance for Sustainable Energy, LLC, for the U.S. Department of Energy (DOE) under Contract No. DE-AC36-08GO28308. Funding provided by the Assessment of Reinforcement Learning for Model NREL Problems Project, funded by the National Renewable Energy Laboratory’s Laboratory Directed Research and Development program. The views expressed in the article do not necessarily represent the views of the DOE or the U.S. Government. The U.S. Government retains and the publisher, by accepting the article for publication, acknowledges that the U.S. Government retains a nonexclusive, paid-up, irrevocable, worldwide license to publish or reproduce the published form of this work, or allow others to do so, for U.S. Government purposes.}
}
    


%



\IEEEtitleabstractindextext{%
\begin{abstract}
Mobility on demand (MoD) systems show great promise in realizing flexible and efficient urban transportation.  However, significant technical challenges arise from operational decision making associated with MoD vehicle dispatch and fleet rebalancing.  For this reason, operators tend to employ simplified algorithms that have  been demonstrated to work well in a particular setting.  To help bridge the gap between novel and existing methods, we propose a modular framework for fleet rebalancing based on model-free reinforcement learning (RL) that can leverage an existing dispatch method to minimize system cost. In particular, by treating dispatch as part of the environment dynamics, a centralized agent can learn to intermittently direct the dispatcher to reposition free vehicles and mitigate against fleet imbalance.  We formulate RL state and action spaces as distributions over a grid partitioning of the operating area, making the framework scalable and avoiding the complexities associated with multiagent RL.  Numerical experiments, using real-world trip and network data, demonstrate that this approach has several distinct advantages over baseline methods including: improved system cost; high degree of adaptability to the selected dispatch method; and the ability to perform scale-invariant transfer learning between problem instances with similar vehicle and request distributions.
\end{abstract}

}

\maketitle

\IEEEdisplaynontitleabstractindextext

%
\IEEEpeerreviewmaketitle

\section*{\textbf{Notation}}
\addcontentsline{toc}{section}{Notation}
\begin{IEEEdescription}[\IEEEusemathlabelsep\IEEEsetlabelwidth{$\textit{orig/dest}$}]
\item[$\mathbb{R}$] Real numbers
\item[$\mathbb{N}$] Natural numbers\\
\item[\textbf{Entities}]
\item[$r$] Passenger request
\item[$v$] Vehicle
\item[$\mathcal{V}$] Vehicle set
\item[$\mathcal{R}$] Trip request set
\item[$\mathcal{R}_t^w$] Set of waiting requests at time $t$
\item[$\mathcal{R}_t^{rebal}$] Set of rebalance requests at time $t$\\
\item[\textbf{Simulation}]
\item[$n_{cap}$] Passenger capacity
\item[$n_{pass}$] Number of passengers
\item[$t_{wait}$] Accumulated request wait time
\item[$\overline{t_{wait}}$] Maximum wait time 
\item[$\Delta t_s$] Simulation time step
\item[$\Delta t_d$] Dispatch interval
\item[$\Delta t_r$] Rebalance interval\\
\item[\textbf{Rebalancing}]
\item[$N_x, N_y$] Grid dimensions
\item[$m, n$] Grid indexes
\item[$\textbf{x}$] Position, $(x,y)$
\item[$G/G^{-1}$] Aggregating/Disaggregating function
\item[$R$] Matrix of aggregated rebalance requests
\item[$V$] Matrix of aggregated free vehicles\\\
\item[\textbf{Reinforcement Learning}]
\item[$\pi$] Policy function
\item[$\theta$] Policy parameter vector
\item[${\alpha}_t$] Action at time $t$
\item[$s_t$] State at time $t$
\item[$\gamma$] Discount factor
\item[$r(s_t, \alpha_t)$] Step reward
\item[$\mathcal{V}_{\pi}(s_t$)] Value function
\item[$\mathcal{Q}_{\pi}(s_t, \alpha_t$)] Action-value function\\
\item[\textbf{Algorithms}]
\item[NR] No rebalance
\item[RR] Random rebalance
\item[SAR] Simple anticipatory rebalance
\item[SAR*] SAR with perfect forecasting
\item[t-SAR] Transferred SAR
\item[RL] Reinforcement learning
\end{IEEEdescription}

\section{Introduction}
%
%
%
%
\IEEEPARstart{T}{he} rapid growth of mobility-on-demand (MoD) transportation services in recent years is reshaping the landscape for mobility patterns of people living in major urban areas. Significant changes in urban mobility have spearheaded new business models, most notably transportation network companies (TNC), such as Uber, Lyft, and DiDi, which pair riders and drivers through the use of mobile phones. TNCs occupy a substantial share of the market in cities across the  United States and abroad. Ride-sharing  generated an estimated \SI{44}[\$]{B} world-wide in 2017 with revenue projected to double through 2025 \cite{oh2020assessing}. The success of these companies can be attributed to many factors, including lower waiting times, moderate cost, and passenger convenience. 

One of the central challenges in operating MoD systems is fleet imbalance, where supply and demand of available vehicles tends to become mismatched geographically over time (see, e.g., \cite{pavone2012robotic}).  To address this, operators can perform fleet rebalancing actions throughout the day in an attempt to move empty vehicles to locations that will better serve future demand and decrease overall energy consumption.  Unfortunately, in most contemporary MoD fleets, rebalancing may be both difficult and expensive to accomplish due to the largely independent nature of individual vehicles and drivers, as well as a shortage of realistic models to optimally compute specific rebalance actions.  However, the emergence of connected, autonomous vehicles (CAV) promises to alleviate some of these difficulties both because such vehicles will be highly amenable to centralized control, and not subject to driver behavior or other related uncertainties.  Thus the fleet rebalancing problem is an increasingly important computational problem in MoD systems.

In this paper, we present a novel framework that uses reinforcement learning (RL) to solve the MoD fleet rebalancing problem by \emph{learning the system dynamics induced by a pre-existing dispatch scheme}. Reinforcement learning is a field of artificial intelligence  with the goal of learning optimal decision policies that maximize a cumulative reward by repeated interaction with a target system or simulation \cite{bertsekas2021multiagent} and is utilized for a wide range of applications including, but not limited to, cyber-physical systems \cite{liu2020parallel} and robot locomotion \cite{cao2019distributed}.  We consider the specific problem of learning to reduce the aggregate customer wait time over an optimization horizon, although the RL algorithm we describe is readily extensible to other, potentially multi-objective, cost functions.  We demonstrate that this framework yields significant improvement for a specific choice of dispatch heuristic by simulating a system with a data-driven network model using both dynamic vehicle speeds and nonlinear travel paths.  Additionally, we show that the use of distributional action and state spaces enables policy transfer between problems instances that have similar spatio-temporal  distributions. Specifically, a model trained on a system with a small number of agents can be readily applied to a large number of agents while retaining much of the control performance, without additional training.  Taken together, we hope that this work can make fleet rebalancing more accessible for operators, particularly those with proprietary dispatch schemes, by allowing the RL agent to learn how best to leverage such schemes in the context of the target system.

The remainder of the paper is organized as follows. In Section \ref{litreview}, we provide a comprehensive review of MoD system control covering both model-based and model-free approaches. In Section \ref{methodology}, we formulate the fleet rebalancing problem and describe the use of both RL and baseline methods for solving it. Section \ref{numericalresults} presents the results of numerical experiments used to validate the RL approach. Finally, Section \ref{conclusions} gives some concluding remarks and proposes potential future work.

\section{Literature Review}\label{litreview}

MoD systems have received significant attention in recent years due to advancements in technology, consumer preferences, and many economic, environmental and societal factors. Recent interest has motivated numerous published works highlighting cost-to-demand benefits \cite{dandl2018comparing}, communication challenges \cite{belakaria2019fog}, as well as optimizing charging patterns of MoD systems with electric vehicles \cite{ammous2018optimal}. Fleet rebalancing studies have been conducted focusing on car rental \cite{bazan2018rebalancing, lu2018optimizing, nourinejad2015vehicle} and public bike-sharing systems \cite{schuijbroek2017inventory, chemla2013bike, chiariotti2018dynamic}. However, since we are interested in rebalancing of ride-sharing vehicle fleets, these studies are beyond the scope of this paper.

At a high level, approaches to solving operational MoD problems may be broadly categorized as (a) model-based and (b) model-free.

\subsection{Model-Based Methods}
We use the term ``model-based'' to describe methods which ascribe an explicit model to system dynamics and use this model to identify optimal decisions. While powerful, model-based methods are usually very complex and become intractable for large-scale systems. An extensive review of recent works regarding model-based solutions to the problem of vehicle relocation issues in vehicle sharing networks can be found in \cite{illgen2019literature}. Numerous studies proposed and developed system models that vary widely and include queuing \cite{zhang2016control}, fluidic \cite{pavone2012robotic}, network flow \cite{rossi2018routing}, and data-driven \cite{lei2020efficient} approaches. Model-based methods can be further categorized into mathematical optimization and simulation-based methods.

Various works have addressed vehicle fleet rebalancing as a complex optimization problem. In \cite{warrington2019two}, the authors approach the problem as a dynamic repositioning and routing problem (DRRP) with stochastic demands. The authors of \cite{zhang2018analysis} propose a rebalancing algorithm using queuing Jackson networks. A scalable model predictive control (MPC) solution approach for fleet rebalancing on AMoD systems was provided in \cite{carron2019scalable}. The authors in \cite{chen2019dynamic} describe a mathematical program of static user equilibrium for an MoD system, and its solution was used as the basis for a linear program built to rebalance empty vehicles.  Another model-based method is the Simple Anticipatory Rebalancing (SAR) method \cite{fagnant2015dynamic,wen2017rebalancing}, which uses a rule-base algorithm to balance instantaneous supply and demand.


\subsection{Model-Free RL}
Explicit control models offer some advantages, in most real-world applications the relevant dynamics are extremely complex and deriving exact models is not possible. Moreover, model-based methods may require strong assumptions and relaxations, e.g. constant arrival rates that are difficult to validate in real-world applications, and may scale poorly, e.g., due to integer constraints. As a result, model-free methods based on RL for addressing vehicle dispatch and rebalancing have started to gain popularity in the literature. The most significant advantage of model-free RL compared to model-based approaches is that a forward simulator suffices to compute a stochastic control policy without relying on an explicit mathematical model of the system dynamics or exogenous factors. Furthermore, the use of deep neural networks makes it possible to evaluate the policy in real-time because computing actions is equivalent to making one forward pass, and this step can be further accelerated via the use of specialized hardware such as a Graphical Processing Unit (GPU). Well known challenges for model-free RL methods include high sample complexity, the need for trade-off between policy exploration and exploitation, problem-dependent reward shaping, and determination of an appropriate neural architecture for the policy network.  As is also the case with model-based methods, differing formulations of the state and action spaces can have a large impact on control performance. 

Previous works on model-free approaches for MoD fleet rebalancing can be categorized into two main classes: i) centralized and ii) decentralized. In the former case, the vehicles are controlled by a centralized agent trained to rebalance them by optimizing a specific objective, i.e. maximizing profits, minimizing travel time. In contrast, decentralized methods allow for each vehicle to be its own agent, trained on either cooperative or competitive manner. These are collectively known as multi-agent RL (MARL) methods.

\subsubsection{Centralized methods}
A recent study in \cite{mao2020dispatch} utilized an actor-critic method to learn an optimal vehicle dispatch policy. The environment was designed as a directed graph, where each node represented a different map zone. However, no rebalancing actions were considered. 
Reference \cite{turan2020dynamic} presented an RL approach for profit maximization and fleet rebalancing cost minimization using a dynamic pricing autonomous MoD framework.  While this framework bears some similarity to our approach, there are significant differences. This RL method focuses on vehicle charging/discharging policies, whereas we approach the problem as a customer waiting time minimization. Furthermore, the action space for the method described in \cite{turan2020dynamic} consists of ride prices and vehicle routing/charging decisions, while our method utilizes a probabilistic action space describing the agent's belief of outstanding customer demand. 
The authors in \cite{gao2018optimize} presented an RL method for addressing the problem of taxi dispatching and rebalancing. Their objective was the maximization of taxi driver long-term profits through an optimal driving strategy using $Q$-learning and sets of discrete states and actions over a grid-shaped map. 
Reference \cite{guo2020deep} proposed a double deep $Q$-learning framework for vehicle routing in a ride-sharing AMoD system where idle vehicles get rebalanced on any given time interval to accommodate future requests.
The method described in \cite{fluri2019learning} used a combination of RL and mixed integer-linear programming for fleet rebalancing and Euclidean partite matching for vehicle dispatching. The authors utilized a hierarchical binary partition of the region of interest and employed tabular $Q$-learning for the RL method. They also compared their method with two well-established rebalancing methods. 
Our method differs significantly from those presented in \cite{gao2018optimize,guo2020deep,fluri2019learning}, especially in terms of action spaces. References \cite{gao2018optimize,guo2020deep} consider simplified, discrete action-spaces denoting single-vehicle actions. The study in \cite{fluri2019learning} initially defined the action as the desired vehicle distribution, however it was later discretized in order to be used in the tabular $Q$-learning framework.

\subsubsection{Decentralized methods}
Reference \cite{al2019deeppool} proposed a vehicle ride-sharing framework using a deep $Q$-network for learning optimal policies of individual vehicles in a distributed, uncoordinated manner. The framework considered passenger satisfaction and vehicle utilization as the main objectives. In \cite{lin2018efficient}, the authors used multi-agent RL, with each available vehicle being an individual agent. Two different RL-based approaches were considered, namely an actor-critic and a deep $Q$-network, and the actions were masked to return only an available subset of actions for each agent. Similar to our method, the state-space was formulated using distributions of available vehicles and requests in grid cells. However, the possible actions were discrete, denoting the grid cell an agent would move on at any given time.
Another example of utilizing multi-agent RL for fleet rebalancing appeared in \cite{gueriau2018samod}. The authors developed a car sharing and dynamic ride-sharing system, where both rebalancing and passenger assignments were learned and executed by individual agents. A $Q$-learning method was utilized. The authors considered also a set of baselines for comparison purposes and employed various metrics extending to both requests and vehicles. A recent study found in \cite{wen2017rebalancing}, the authors developed a deep $Q$-learning approach for rebalancing idle vehicles. Although this study has few similarities with ours, it differs in the discrete, agent-level nature of the action space.  The authors compared their proposed method against two model-based approaches, namely an optimal and a simple anticipatory rebalancing strategy.

\subsection{Main Contributions of this Paper}
Our proposed method addresses the difficulty in formulating and solving model-based optimization problems for MoD control by adopting a model-free RL formulation.  The method is fundamentally centralized and uses a novel, distributional representation for state and action spaces to direct a pre-existing dispatch algorithm for trip assignment and routing of individual vehicles.  In doing so, the centralized problem remains tractable for large systems while, at the same time, avoids the intrinsic difficulties of multi-agent RL arising from non-stationary learning, partial observability, and credit assignment.  Additionally, learned policies can be readily transferred to new problem instances with similar spatio-temporal distributions.

We summarize the main contributions of this paper as:
\begin{enumerate}
\item \emph{Modularity}.  The proposed RL agent is designed to wrap an existing algorithm for trip assignment and routing, and directly leverage it to execute rebalancing actions via the creation of rebalancing requests.  This modularity enables a user to take advantage of proprietary algorithms that may embody hard-earned heuristics about the system and its operation, and significantly lowers the technical bar for bringing RL to bear on the rebalancing problem.
\item \emph{Distributional Actions and Policy Transference}. The action space in our approach is modeled as a spatial distribution, a fact that enables policies to be transferred between scenarios with vastly different scales so long as their underlying request distributions are similar. We demonstrate, for example, that a policy learned on a simulated system with 1,500 requests and 100 vehicles can be effectively transferred -- with no additional training -- to a system with 100x the number of entities.
\item \emph{Data-Driven Vehicle Routing Network}.  We demonstrate the efficacy of our approach on a simulated system in which both requests and network properties are derived from real-world data.  In particular, request data are sampled from the Chicago TNP data set and an intermediate road network with spatially varying speeds is derived from TomTom data.  This second fact takes us beyond the state-of-the-art in most papers, in which vehicle travel is typically modeled as point-to-point along straight line segments with globally averaged speed, and provides additional evidence that our approach could be successfully applied to real-world systems.
\end{enumerate}

\section{Methodology}\label{methodology}

\newcommand{\rlstate}[1]{s_{#1}}
\newcommand{\rlaction}[1]{a_{#1}}
\newcommand{\vehs}[0]{\mathcal{V}}
\newcommand{\reqs}[0]{\mathcal{R}}
\newcommand{\tsar}[0]{t-SAR\,}
\newcommand{\sarstar}[0]{SAR$^*\,$}

Preliminaries including notation and a high level overview of model-free RL are provided in Section \ref{ssec:preliminaries}. In Section \ref{ssec:dispatch} we formulate the dispatch problem, and then in \ref{sssec:rebalance} describe how fleet rebalancing can be achieved using any heuristic solution method for the dispatch problem.  Section \ref{ssec:baselines} describes baseline rebalancing methods, while \ref{ssec:reb_rl} formulates the use of RL for this problem.

\subsection{Preliminaries}\label{ssec:preliminaries}

\subsubsection{Notation}\label{sssec:notation} We adopt a notational convention that makes it simple to describe local, aggregate, and time dependent properties of  entities in an MoD system.  First, we denote a local property $z$ of a generic entity $e$ by $e(z)$; for example, $e(st)$ to denote its scalar status and $e(\mathbf{x})$ for vector position.  We use a subscript $t$ to denote time dependent properties as in $e(st_t)$ or $e(\mathbf{x}_t)$.  The aggregate set of entities of a certain type is then straightforward, e.g., $\{e: e(\mathbf{x}_t) = \mathbf{x}'\}$ would denote the set of all entities whose position equaled $\mathbf{x}'$ at time $t$.

We consider the operation of a fleet of capacitated vehicles, $\vehs=\{v_i\}_{i=1}^{|\vehs|}$, with the state of each vehicle given by its
\begin{itemize}[leftmargin=*]
    \item \emph{status}: $v(st)\in\{f,o\}$ where $f$ means ``free" or available to be dispatched; $o$ means ``occupied" or unavailable to be dispatched.
    \item \emph{position}: $v(\mathbf{x})$ with  $xy$-coordinates, $ (v(x), v(y))$; 
    \item \emph{capacity}: $v(n_{cap})$, maximum number of passengers.
\end{itemize}
\noindent Similarly, we consider a set of trip requests, $\reqs=\{r_i\}_{i=1}^{|\reqs|}$, with the state of each request given by its
\begin{itemize}[leftmargin=*]
    \item \emph{status}: $r(st)\in\{w,a,o,d,f\}$ where $w$ means ``waiting" for vehicle assignment; $a$ means ``assigned" to vehicle but not yet picked up; $o$ means ``occupying" a vehicle; $d$ means ``delivered" to destination; $f$ means ``failed", i.e., request was not assigned to a vehicle before exceeding its maximum waiting time;
    \item \emph{position}: $r(\mathbf{x}) = (r(x), r(y))$. We further differentiate between origin and destination positions by $r(\mathbf{x}_{orig})$ and $r(\mathbf{x}_{dest})$, respectively;
    \item \emph{accumulated wait time}: $r(t_{wait})$, the current total time that a request has waited for a vehicle to be assigned;
    \item \emph{maximum wait time}:  $r(\overline{t_{wait}})$, the maximum wait time before the request is considered ``failed" and removed from the dispatch queue;
    \item \emph{number of passengers}: $r(n_{pass})$.
\end{itemize}
\noindent With this notation we can easily describe subsets of entities needed to discuss dispatch and rebalance operations, for example, $\{r: r(st_t) \in \{a,o\}\}$ to denote the set of requests that are either assigned or in transit at time $t$.

\begin{figure}[t!]
  \begin{center}
  \includegraphics[width=\linewidth]{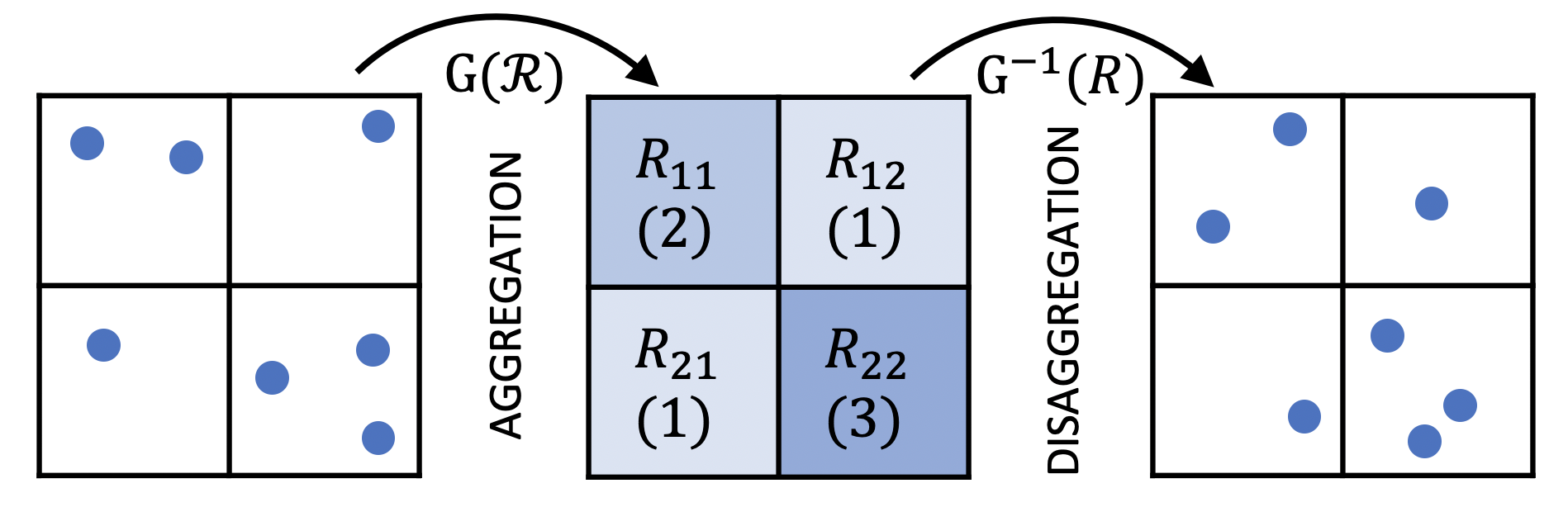}
  \caption{\footnotesize Illustration of the aggregation and disaggregation mappings needed to implement the distributional RL state and action spaces on a $2\times2$ grid.  The elements of matrix $R$, representing the number of requests in each grid cell, are indicated in parentheses.  The RL state matrices $R_t$ and $V_t$ are generated via aggregation, while RL rebalancing actions, in the form of matrices $R_t^{rebal}$, are disaggregated before being fed to the dispatcher.}
  \label{img:distribution-operators}
  \end{center}
\end{figure}

\subsubsection{Entity Aggregation and Disaggregation}\label{sssec:aggregation}
As will become clear below, our approach requires us to be able to go back and forth between entity-level (disaggregate) and distributional (aggregate) views of the system at a given time, as illustrated in Fig. \ref{img:distribution-operators}.  Specifically, we consider partitioning the operating area into a regular $N_x \times N_y$ grid and define a grid mapping, $g: \mathbb{R}^2 \rightarrow \mathbb{N}^2$, that maps $xy-$coordinates of an entity to their cell indexes.  This mapping defines another entity-level property of \emph{grid position}, for example, 
\begin{align}
    r(g(\mathbf{x}))=(r(m), r(n)),
\end{align}
for integers $m\in\{1,\dots,N_x\}, n\in\{1,\dots,N_y\}$, to denote the grid index of a request $r$.  We also define an aggregating function, $G: \mathcal{S} \rightarrow \mathbb{R}^{N_x\times N_y}$,  that counts the number of elements in each grid cell for a given input set, $\mathcal{S}$. For instance, letting $\reqs_t^w$ denote the set of ``waiting" requests at time $t$, 
\begin{align}
    \left[G(\reqs_t^w)\right]_{mn} = \left|\{ r \in \reqs_t^w :  r(g(\mathbf{x})) = (m, n)\}\right|,
\end{align}
denotes the number of such requests in cell $(m,n)$.  Finally, with a slight abuse of notation, we denote a disaggregation operator as $G^{-1}: \mathbb{R}^{N_x \times N_y} \rightarrow \mathcal{S}$ that ``inverts" the aggregation effected by $G$.  In particular, the operation $G^{-1}(G(\reqs_t^w))$ generates a set of $|\reqs_t^w|$ requests with the same grid positions as those in the originating set but, notably, not necessarily the same $xy$-coordinates.  In this paper, we utilize a disaggregator operation that generates positions uniformly at random within each grid cell.  Fig. \ref{img:distribution-operators} illustrates both  aggregation and randomized disaggregation.

\emph{Simple example.} Consider a $1\times2$ grid with cell width and height of 1, and with a single request in grid cell $(1,1)$ at position $(0.2, 0.8)$ and no requests in cell $(1, 2)$.  The aggregation operator simply counts the requests in each cell, such that $[G(\reqs)]_{11}=1$ and $[G(\reqs)]_{12}=0$.  One possible outcome of disaggregation, which preserves cell count but generates random positions, would be a single request at position $(0.5, 0.1)$ in cell $(1, 1)$.

\subsubsection{Time stepping} For ease of exposition, we assume that physical (or simulated) time steps, dispatch steps, and rebalance steps -- denoted by $\Delta t_s$, $\Delta t_d$, and $\Delta t_r$, respectively -- are integer multiples of one another with  $\Delta t_s \leq \Delta t_d \leq \Delta t_r$.    In particular, we constrain the dispatch interval to be an integer multiple of the simulator step, $\Delta t_d = n_d \Delta t_s$, and the rebalance interval to be an integer multiple of the dispatch interval, $\Delta t_r = n_r \Delta t_d$, where $n_d, n_r$ are positive integers. This implies that physical and rebalance steps are similarly related, with $\Delta t_r = n_r n_d \Delta t_s$. While these constraints are not strictly required, they greatly simplify the notation needed to discuss simulation, dispatch and rebalance events that occur on differing time scales.

\subsubsection{Model-Free Reinforcement Learning} \label{ssec:model-free-rl} We next provide a brief overview of model-free RL, referring the reader to standard text such as \cite{sutton2018reinforcement} for further details.  At a high level, the goal in model-free RL is to identify a sequence of actions that maximize the expected discounted future rewards, or \emph{reward-to-go}, obtained by an agent starting from some initial state.  A distinguishing characterisic of model-free RL is that the information needed to identify such a sequence is acquired via trial-and-error interaction with the target system.  Many RL algorithms approximate a \emph{policy} function, denoted by $\pi$, that directly maps states to actions, $\rlaction{t}=\pi(\rlstate{t})$.  An optimal policy, $\pi^*$, is one that generates actions that maximize the reward-to-go, for instance,
\begin{align}
\pi^*=\arg\max_{\pi}\mathbb{E}_{\pi}\left[\sum_{t=0}^{\infty}\gamma^t r(\rlstate{t},\rlaction{t})\right].
\end{align}
Here, $\gamma\in[0, 1]$ is the discount factor, $r(\rlstate{t},\rlaction{t})$ is a step reward, and $\mathbb{E}_\pi(\cdot)$ denotes the expected value of the argument achieved by following the policy $\pi$ over all future time steps.  A common assumption in many modern RL algorithms is that $\pi$ is parameterized by a vector $\theta$, for example, the weights of a neural network.  We denote this dependence by $\pi_\theta$.  For parameterized policies, the above optimization problem can then be recast as
\begin{align}
\theta^*=\arg\max_{\theta}\mathbb{E}_{\pi_\theta}\left[\sum_{t=0}^{\infty}\gamma^t r(\rlstate{t},\rlaction{t})\right].
\label{eq-max-theta-rl}
\end{align}
Because an RL agent's goal is to maximize its reward-to-go, many RL algorithms directly model the \emph{value function},
\begin{align}
\mathcal{V}_{\pi}(\rlstate{t})=\mathbb{E}_{\pi}\Bigg[\sum_{t'=t}^{\infty}\gamma^{t'-t} r(\rlstate{t'}, \rlaction{t'})\Bigg]
\end{align}
which returns the reward-to-go for a given state, attained by following the policy $\pi$ from that state.  The Bellman equation \cite{sutton2018reinforcement},
\begin{align}
\mathcal{Q}_{\pi}(\rlstate{t}, \rlaction{t}) = r(\rlstate{t}, \rlaction{t}) + \mathcal{V}_{\pi}(\rlstate{t+1}),
\end{align}
relates the single-step return for taking action $\rlaction{t}$ in state $\rlstate{t}$ to the value of following policy $\pi$ thereafter.  This expression provides the starting point for many RL algorithms because it provides a value-based approach to defining an optimal policy, namely, by choosing the action that maximizes $Q$. As is the case with the policy function $\pi$, both the value and $Q$-functions may also be parameterized.

In general, RL algorithms typically use some combination of policy and value approximation to learn how to best control the target environment.  Depending on which functions are being approximated, this leads to three broad categories of algorithms: a) value function-based, b) policy gradient-based, and c) actor-critic methods that use both policy (actor) networks for generating actions, and a value (critic) networks to estimate the value of a given state. The algorithm used in this paper, Proximal Policy Optimization \cite{schulman2017proximal}, is a neural network-parameterized actor-critic method that can handle continuous action and state spaces; see Section \ref{rl_implementation} for further discussion.

\subsection{The dispatch problem}\label{ssec:dispatch}

In this paper, we define the \emph{dispatch problem} as one of computing the assignment and routing of free vehicles to service waiting requests (including both pick-up and drop-off), in order to minimize one or more operational costs summarized by a function $f$ of the system state, $\rlstate{t}$.  In particular, we consider solving this problem over a horizon of length $T=N_s \Delta t_s$, stated as
\begin{align}
    \min_{\mathbf{A}, \mathbf{d}} &\sum_{t=0}^{T} f(\rlstate{t} | \mathbf{A}, \mathbf{d}),
    \label{eq-global-min-wait-time}
\end{align}
where $\mathbf{A}=\{A_t: t \mod \Delta t_d = 0\}$ is a sequence of trip assignment matrices and $\mathbf{d}=\{d_t: t \mod \Delta t_d = 0\}$ is a sequence of functions that return associated routing decisions at each dispatch step.  Specifically, $[A_t]_{ij}=1$ if vehicle $v_i$ is assigned to request $r_j$ at dispatch time $t$, and 0 otherwise.  The dependence of $f$ on $\mathbf{A},\mathbf{d}$ is implicit because the assignment and routing decision affect the system state, but are not assumed to be explicit variables in the cost expression.  

Despite its simple statement, Eq. \eqref{eq-global-min-wait-time} can be very challenging to solve: this formulation hides all complex constraints needed to represent the system dynamics, e.g., binary variables for trip assignment, and integer constraints for vehicle capacity.  Furthermore, the optimization is defined over a horizon that,  for real-world systems, encompasses uncertainty from model error, forecasting error, and other exogenous sources.  The inclusion of routing decisions as part of the dispatch solution further increases the complexity.

\subsection{Free vehicle repositioning via dispatch of rebalancing requests}\label{sssec:rebalance} Our rebalancing approach assumes that a heuristic for the dispatch problem in Eq. \eqref{eq-global-min-wait-time} exists and that it outputs a solution at every dispatch step. The exact details of the heuristic (e.g., the use of a lookahead model) are irrelevant; we assume only that it outputs vehicle assignments for waiting requests and associated routing decisions. This solution method can be directly leveraged to reposition vehicles via the introduction of \emph{rebalancing} requests, each of which has no real passengers ($n_{pass}=0$) and identical origin and destination ($\mathbf{x}_{orig}=\mathbf{x}_{dest}$).  When the dispatcher receives a rebalancing request, it assigns and routes a free vehicle to the new location such that the vehicle again becomes freed as soon as it arrives.  Formulated this way, repositioning decisions amount to computing the number and position of rebalancing requests at each rebalance step.

Exploiting a heuristic for Eq. \eqref{eq-global-min-wait-time} to solve the rebalancing problem, then, amounts to computing a ``good" set of rebalancing requests, $\reqs_t^{rebal}$, that strategically repositions free vehicles:  once this set is computed, it can simply be treated as a new cohort of realized trip requests that the dispatcher attempts to service.  While there are many possible ways to generate such a request set, we adopt a grid partitioned view of the system such that rebalancing decisions are cast as discrete distributions over the grid.  These decisions are then disaggregated into rebalancing requests sets using the operations described in Section \ref{sssec:notation}.  In adopting this representation, we are able to learn a scalable, centralized control policy that is readily transferred to problem instances with similar spatio-temporal request distributions, a claim that we support with numerical evidence in Section \ref{sec:res_disc}.  Furthermore, we avoid having to deal with spaces of changing dimension (i.e., differing numbers of entities) which can be challenging for RL-based methods.

\begin{figure}[t!]
  \centering
  \includegraphics[width=\linewidth]{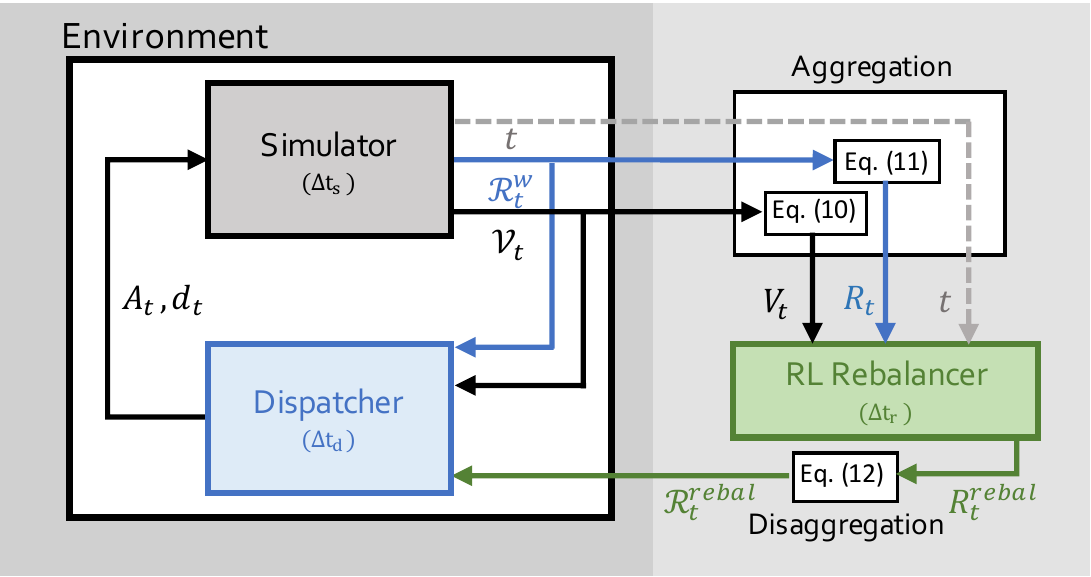}
  \caption{\footnotesize Schematic of the simulation environment with interleaving dispatch and rebalance steps.  The left panel encompasses a simulation environment with a built-in (potentially black-box) dispatch scheme running every $\Delta t_d$ minutes and feeding assignment and routing decisions ($A^t,d^t$ in Eq. (\ref{eq-global-min-wait-time})) into the simulator.  The right panel illustrates how information from the environment is aggregated in order to create the state vector, $\rlstate{t}=(V_t, R_t, t)$, used by RL to create rebalancing actions, $\rlaction{t}=R_t^{rebal}$ in Eq. (\ref{eq-rebalancing-over-G}).  These distributional actions are then disaggregated into a rebalancing request set, $\reqs_t^{rebal}=G^{-1}(R_t^{rebal})$, used by the dispatcher to affect repositioning of free vehicles.}
  \label{img:framework}
\end{figure}

More formally, rebalancing in our framework amounts to finding a sequence of rebalancing request matrices, $\mathbf{R}$, to solve,
\begin{align}
    \min_{\mathbf{R}} & \,\,J_{A,d}(\mathbf{R})\nonumber\\
    \mathbf{R} &=\{R_t^{rebal}:  t = 0, \Delta t_r, \dots, (N_s/n_r)\Delta t_r\},
    \label{eq-rebalancing-over-G}
\end{align}
where $J_{A,d}$ denotes an optimal objective function value for the dispatch problem shown in Eq. \eqref{eq-global-min-wait-time}.  Note that the decision variables coincide with aggregate distributions of rebalancing requests, with the implication that these must be disaggregated before being sent to the dispatcher (see Section \ref{sssec:aggregation}).  Although not obvious at first sight, Eq. (\ref{eq-rebalancing-over-G}) actually represents a bi-level optimization problem because rebalance and dispatch solutions are computed independently, and yet each affects the free vehicle distribution used as input for the other.  This type of problem can be exceedingly difficult to solve, a fact that further justifies the use of a model-free solution method in this context. Fig. \ref{img:framework} presents an overview of the proposed framework.

\subsection{Fleet Rebalancing with Baseline Methods}\label{ssec:baselines}

To assess the value of applying RL for fleet rebalancing, we compare it with three baselines:  a null policy of no rebalance (NR), random rebalance (RR), and simple anticipatory rebalance (SAR). 

\subsubsection{No Rebalance} This baseline method simply uses the black-box dispatch algorithm to solve Eq. \eqref{eq-global-min-wait-time} for each control time step.  No lookahead model is used and no rebalancing requests are created:  the fleet distribution simply evolves according to the dynamics arising from repeatedly solving the dispatch problem and executing the subsequent commands.

\subsubsection{Random Rebalance} We include a purely random method to illustrate that the act of rebalancing can be detrimental (worse than \textit{no rebalance}) if performed poorly. At each rebalance time step, this baseline method generates a uniform random distribution of rebalancing requests.  The size of the request set can range from 0 to $N_{rebal}$ requests.

\subsubsection{Simple Anticipatory Rebalance}\label{sssec:sarstar} This model-based method is similar, albeit slightly modified, to the one used as a benchmark in \cite{wen2017rebalancing}. The idea behind the SAR algorithm is to use a forecast of the request distribution at each time step and apply a rebalancing action that anticipates it.  In this paper, we consider a specific implementation of this algorithm, which we denote by \sarstar, that assumes perfect knowledge of the request distribution over the next rebalance step.  In particular, at each rebalance step, the set of rebalancing requests is set equal to the realized request distribution over the next rebalance interval:
\begin{align}
    \text{SAR}^*: \quad \reqs_t^{rebal} = \bigcup_{t'} \reqs_{t'}^w,
    \label{eq:sar-star-rebalance}
\end{align}
where $t'\in\{t, t+\Delta t_s,\dots,t+\Delta t_r\}$ denotes all time steps in the upcoming rebalancing step. The dispatch heuristic for Eq. \eqref{eq-global-min-wait-time} is then applied over the horizon $[t', t'+\Delta t_r]$.  While \sarstar is not provably optimal, its use of perfect forecasting leads to far better performance over the other baseline methods (Section \ref{numericalresults}), making it a meaningful benchmark for comparison with RL.

\subsection{Fleet Rebalancing with Reinforcement Learning}\label{ssec:reb_rl}

In this paper, we consider a model-free RL formulation for solving Eq. \eqref{eq-rebalancing-over-G} which is  re-cast as a reward maximization problem by simply negating the objective.  The overarching framework is illustrated in Fig. \ref{img:framework}.  We define an observation as $\rlstate{t}=(V_t, R_t, t)$, where $V_t$ represents the aggregate free vehicle count at time $t$, and $R_t$ represents the aggregate waiting request count over the previous rebalance interval.  More precisely, $V_t$ and $R_t$ are $N_x \times N_y$ matrices with elements
\begin{align}
    [V_t]_{mn} =& \left|\{v: v(st_t) = f, \,  v(g(\mathbf{x}_t))=(m,n)\}\right|,
    \label{def-vehicle-dist}\\
    [R_t]_{mn} =& \left|\bigcup_{t'} \{r: r(st_{t'}) = w,  \, r(g(\mathbf{x}_{t'}))=(m,n)\}\right|,
    \label{def-request-dist}
\end{align}
where $t'\in\{t-\Delta t_r, t-\Delta t_r + t_s,\dots, t\}$ denotes all time steps from the preceding rebalance interval. We included time, $t$, in the state due to the diurnal nature of the data for which time of day is a highly predictive feature.  

Just as the decision variables in Eq. \eqref{eq-rebalancing-over-G} are the rebalancing request distributions at each time step, we define the RL actions to be precisely those decisions, i.e., we take $\rlaction{t}=R_t^{rebal}$.  Specifically, $\rlaction{t}$ is an $N_x \times N_y$ matrix that disaggregates into rebalancing request sets for each grid cell $(m,n)$ that are fed to the dispatch algorithm to effect repositioning,
\begin{align}
    \text{RL:} \quad \reqs_t^{rebal} = G^{-1}(R^{rebal}_t).
    \label{eq-disaggregation}
\end{align}
As is always the case with model-free RL, Eq. \eqref{eq-rebalancing-over-G} is solved sequentially: an action is generated and applied at each rebalance time, depending on the realized state.

\begin{algorithm}[t!]
 \caption{Time evolution of a simulated MoD system with interleaved dispatch and rebalancing}
 \begin{algorithmic}[]\label{alg:simloop}
  \FOR{$t = 0, \Delta t_s, \dots, N_s \Delta t_s$}
    \IF{$t \,\, \text{mod}\,\, \Delta t_d = 0$}
    \STATE{- Dispatch: Assign and route free vehicles to waiting requests, $\reqs_t^w$, using any solution method for Eq. \eqref{eq-global-min-wait-time}}.
    \ENDIF
    \IF{$t \,\, \text{mod} \,\, \Delta t_r = 0$}
      \STATE{- Generate rebalancing requests, $\reqs_t^{rebal}$, e.g., using Eq. \eqref{eq:sar-star-rebalance} for \sarstar or Eq. \eqref{eq-disaggregation} for RL.}
      \STATE{- Dispatch: Assign and route free vehicles to rebalancing requests, $\reqs_t^{rebal}$, using any solution method for Eq. \eqref{eq-global-min-wait-time}.}
    \ENDIF 
  \ENDFOR
 \end{algorithmic} 
\end{algorithm}

\section{Numerical Studies}\label{numericalresults}

\subsection{Simulation}
\subsubsection{Agent-based simulator}\label{sec-simulator}
We implemented an agent-based simulator that enabled us to study the impacts of different rebalancing schemes on the system objective.  Algorithm \ref{alg:simloop} summarizes the interleaving of dispatch and rebalance steps in simulation. We used one-minute time steps ($\Delta t_s=1$) and dispatch steps ($\Delta t_d=1$) and 60-minute rebalance steps ($\Delta t_r=60$), over simulated episodes of length 24 hours ($N_s = 1440$) starting at midnight.  Importantly, vehicle movement in our simulator was not modeled as point-to-point between origin and destination; rather, vehicles move along network edges with varying speeds as estimated from a TomTom data set described in Section \ref{sec-tomtom}.  Vehicle positions were initialized uniformly at random over the operating area, and all vehicles were assumed to have four-passenger capacity ($n_{cap}=4$) and a status of free ($v(st)=f$) at the beginning of the simulation.  While we hope to include refueling dynamics in future work, for the purpose of this study we did not include refueling details (vehicles are not required to refuel).  Request origins, destinations, and activation times were taken directly from the Chicago TNC data set and were assumed to expire after $\overline{t_{wait}}=30$ minutes, in which case the request was considered ``failed".  Aggregation was performed using a $5\times5$ grid ($N_x=N_y=5$) as shown in Fig. \ref{img:maps}; random disaggregation as described in Section \ref{sssec:aggregation} was used to create rebalancing requests from the RL action matrix, $\rlaction{t}=R_t^{rebal}$.

\begin{figure}[t!]
\begin{centering}
\begin{subfigure}{.18\linewidth}
  \centering
  \includegraphics[width=\linewidth]{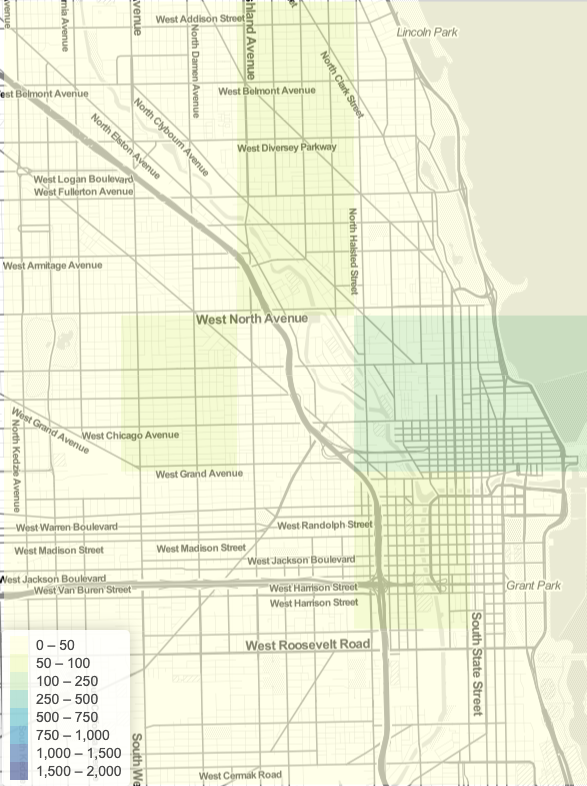}
  \caption{}
  \label{fig:sfig1}
\end{subfigure}
\begin{subfigure}{.18\linewidth}
  \centering
  \includegraphics[width=\linewidth]{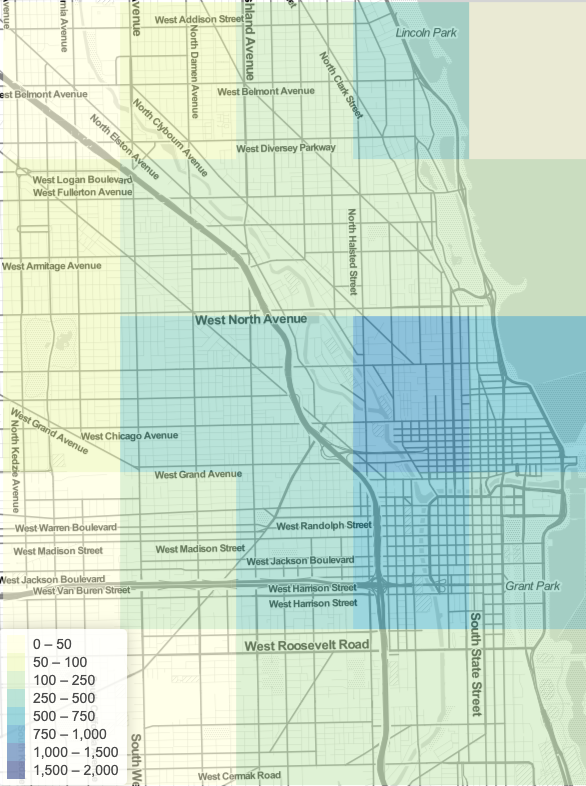}
  \caption{}
  \label{fig:sfig2}
\end{subfigure}
\begin{subfigure}{.18\linewidth}
  \centering
  \includegraphics[width=\linewidth]{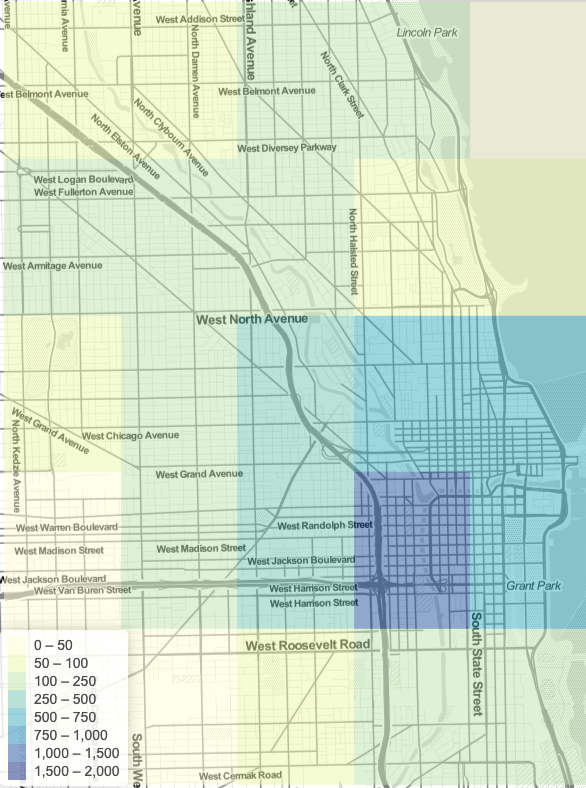}
  \caption{}
  \label{fig:sfig3}
\end{subfigure}
\begin{subfigure}{.18\linewidth}
  \centering
  \includegraphics[width=\linewidth]{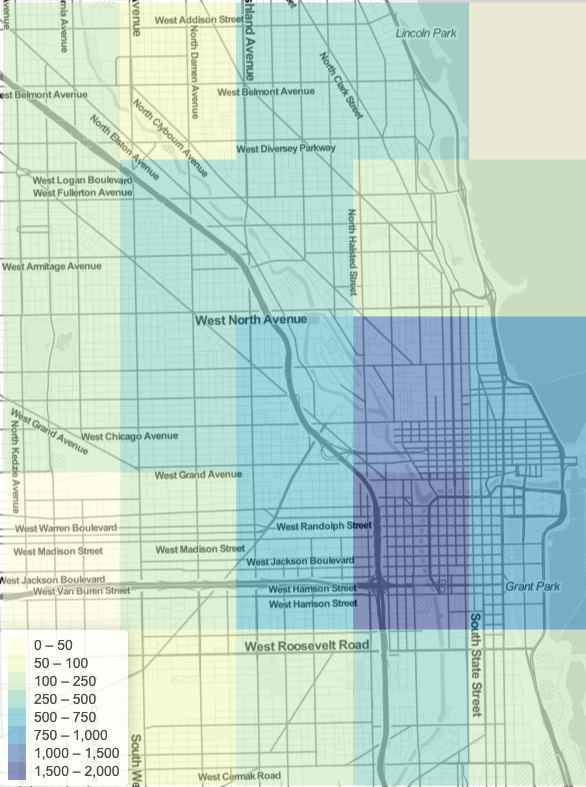}
  \caption{}
  \label{fig:sfig4}
\end{subfigure}
\begin{subfigure}{.18\linewidth}
  \centering
  \includegraphics[width=\linewidth]{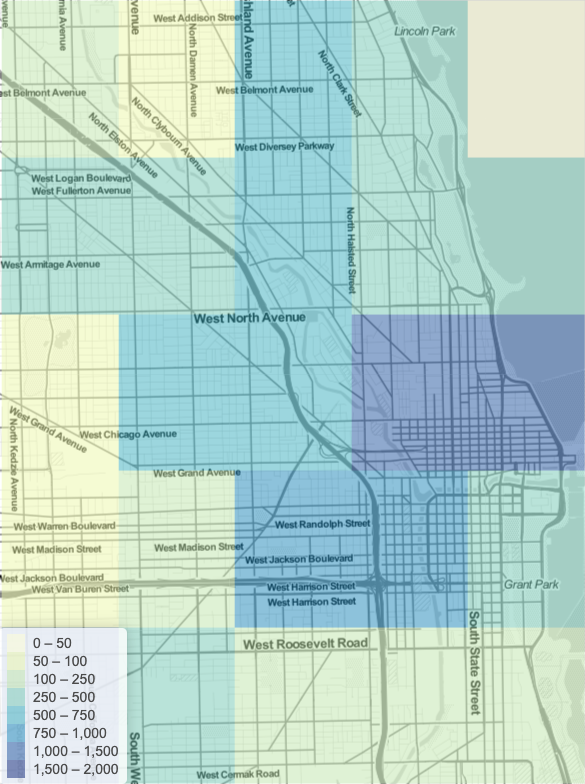}
  \caption{}
  \label{fig:sfig5}
\end{subfigure}
\caption{\footnotesize Spatial distributions of hourly trip demand on the date used for training, Friday, September 13, 2019, at (a) 01:00 (b) 07:00 (c) 12:00 (d) 17:00 (e) 22:00.  Higher demand is indicated by darker colors.}\label{img:sptial_dist}
\label{fig:fig}
\end{centering}
\end{figure}

\subsubsection{Dispatch model}\label{sec-dispatch-model} For our studies, we used a simple, greedy heuristic  to solve Eq. \eqref{eq-global-min-wait-time}.  Specifically, for each waiting request, $r^w_j$, we  assign the nearest free vehicle with sufficient capacity by solving,
\begin{align}
\min_{i} & \quad \left|\left| v_i(\mathbf{x}) - r^w_j(\mathbf{x}_{orig}) \right|\right|_2 &\\
\text{s.t.} & \quad v_i(st) = f, & \text{(free status)} \nonumber\\
            & \quad v_i(n_{cap}) \geq r_j(n_{pass}). & \text{(capacity)} \nonumber
\end{align}
Routing was determined using yet another heuristic described in the next section.  Performance of this dispatch and routing scheme is intentionally suboptimal:  by treating the dispatch scheme as a black-box part of the environment, a planning agent can learn the induced dynamics and translate this learning into improved performance using standard model-free RL.

\subsubsection{Vehicle routing and transportation network model}\label{sec-tomtom}
\begin{figure}[t!]
  \includegraphics[width=\linewidth]{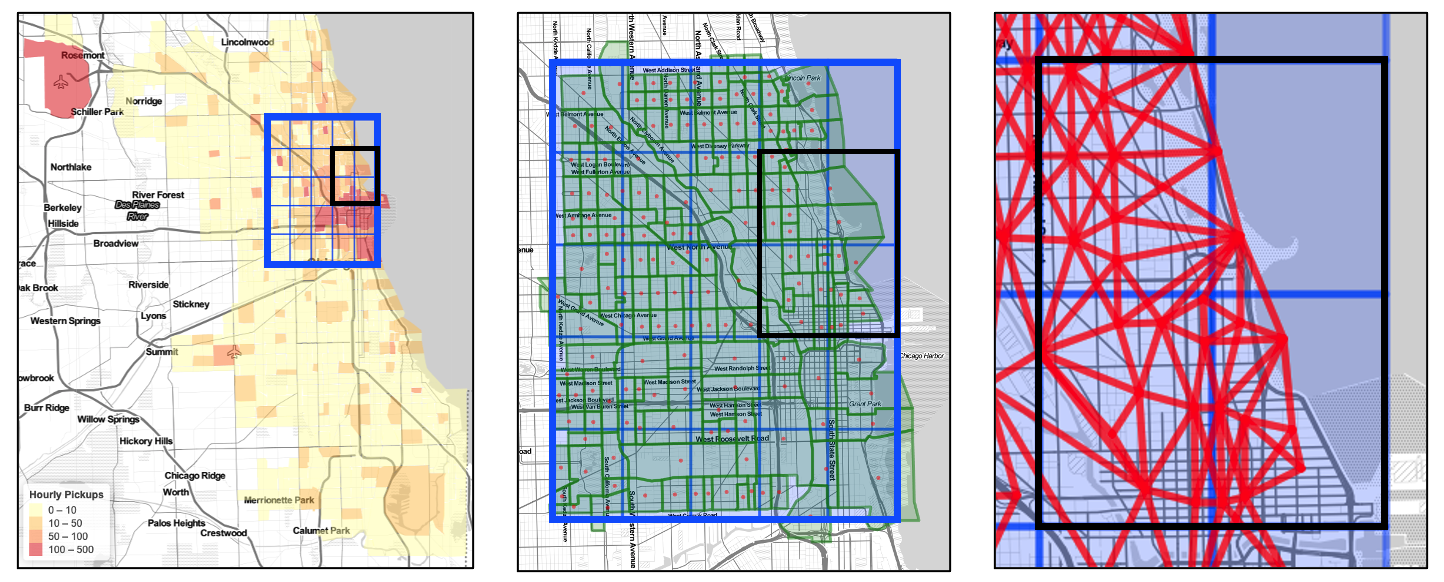}
  \caption{\footnotesize Map of Chicago showing the operating area considered in this paper, at greater resolution going from left to right.  The aggregation grid is illustrated in blue, census tract boundaries in green, and census tract centroids and associated network edges in red.  The routing network consists of edges (red lines, right panel) connecting the centroid centers (red dots, center panel) with tracts that share boundaries (green regions, center panel).  The utilization of census tract centroids is motivated by the structure of data set, as described in Section \ref{ssec:case-studies}.}\label{img:maps}
\end{figure}

As mentioned above, in our studies we adopt a simple heuristic for routing as a post-processing step to trip assignment.  Since the raw trip data is aggregated at census tract level (discussed in greater detail in Section \ref{ssec:case-studies} below), we assume all trips start and end at centroids of census tracts, ignoring detailed origin and destination location information in the simulation. Then, a sketch vehicle routing network is developed with centroids of census tract as the network nodes using real world traffic data obtained from TomTom, a third party traffic data provider \cite{tomtom}. The connectivity between the network nodes is determined by the adjacency of census tracts. If two census tracts are adjacent to each other, i.e., share a common boundary, the centroids of the two census tracts are connected with a link. The weights of links are the average vehicle travel time during the day between the linked two nodes to be simulated. Links are directional in the network as vehicle travel time from location A to location B can vary from travel time from location B to location A depending on the traffic. Based on the rules described above, an adjacency matrix was developed using travel time data queried from TomTom. The vehicle routing network is then derived from the adjacency matrix as displayed in Fig. \ref{img:maps}, and Dijkstra's algorithm \cite{dijkstra1959note} is used to compute the shortest path (in travel time) between origin and destination as the values of network weights (travel time) are non-negative. Dijkstra’s algorithm searches the shortest path for any weighted directed graph with non-negative weights by analyzing the distances between source node to all other nodes in a graph. The algorithm caches the currently known shortest distance from the source node to each node and updates the values if a shorter path is found. It works on graphs with cycles as long as they are positive weight cycles. Negative weight cycles produce incorrect results since they cause cycling an infinite number of times to minimize the total distance.

\subsection{Optimization Objective}\label{ssec:optimization-objective}

In this paper, we consider the total customer wait time,
\begin{align}
    f(\rlstate{t}) = \sum_{r\in\reqs_w^t} r\left(n_{pass}\right) \Delta t_s,
    \label{eq-f-total-wait-time}
\end{align}
in Eq. \eqref{eq-global-min-wait-time}, where $\reqs^w_t=\{r: r(st_t)=w\}$ is the set of requests waiting for vehicle assignment at time $t$.  This objective function is a reasonable proxy for quality of service of the fleet and one that we expect to be strongly related to fleet imbalance: intuitively, wait times will increase when fewer free vehicles are properly positioned to anticipate demand.  We note that Eq. \eqref{eq-f-total-wait-time} is easily extended to include other costs such as total empty vehicle miles traveled or fuel costs.  We hope to extend this study in the future to consider these and other multi-objective problems.

\begin{figure*}[t!]
\centering
  \includegraphics[width=.8\textwidth]{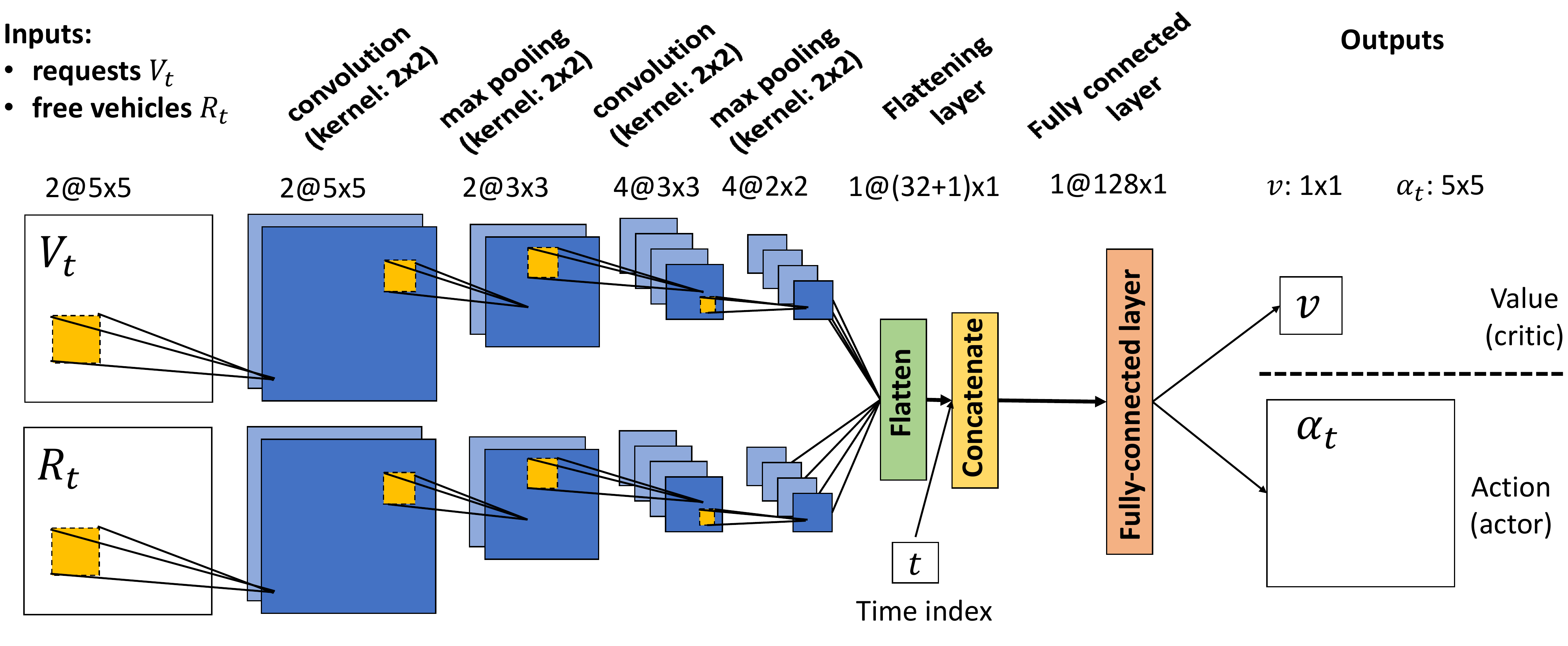}
  \caption{\footnotesize Actor-critic networks used in PPO. Array inputs are filtered through two convolutional layers with same padding, followed by max-pooling layers that are then flattened and concatenated. A normalized time index is fed to the network after flattening.}\label{img:networks}
\end{figure*}

\subsection{RL Implementation Details}\label{rl_implementation}
In this paper, we use the Proximal Policy Optimization (PPO) algorithm \cite{schulman2017proximal} to solve the rebalancing problem.  PPO is a popular, deep learning-based algorithm for environments with continuous action and state spaces and known to work well off-the-shelf for many problems.  It is important to note that our approach does not make any assumptions about the RL algorithm itself, other than that it can handle both continuous action and state spaces.

In order to handle the distributional state space, $\rlstate{t}=(V_t, R_t, t)$, and action space, $\rlaction{t}=\pi(\rlstate{t})$, described in Section \ref{sssec:notation}, we developed a neural network architecture for PPO based on convolutional layers as illustrated in Fig. \ref{img:networks}.  This choice was motivated by an interpretation of the aggregation matrices, $V_t$ and $R_t$, as pixelated images such as those shown in Fig. \ref{img:sptial_dist}, and the fact that convolutional layers are known to excel in image-based tasks (see, e.g., \cite{iandola2016squeezenet}).  Although we omit details here, we observed in our experiments that the use of convolutional layers significantly boosted performance over fully connected layers with a comparable number of trainable parameters.

All experiments were conducted on NREL's supercomputer, Eagle, using a single CPU node with 36 cores.  Training time was limited to 40 wall-clock hours, after which the policy checkpoint with the highest episode reward was selected for future experimentation. For scaling up RL training to multiple processors, we used the open-source library RLlib \cite{rllib}, which enabled us to utilize all 36 cores to parallelize episode rollouts and experience collection.  In our experiments, we found that the most significant factor impacting policy performance was the neural architecture described in the previous paragraph; all other hyperparameters in RLlib's PPO implementation were set to the default values.

\subsection{Case Studies}\label{ssec:case-studies}

We utilized the proposed framework for deriving rebalancing policies using publicly available real-world ride-hailing data shared by the City of Chicago \cite{tnp_source}. This data, hosted in the Chicago Transportation Network Provider (TNP) portal \cite{chicago_tnp_data}, consists of aggregated information provided by the various TNCs operating within Chicago (notably including Uber and Lyft). The Chicago TNP data consists of three separate tables: a trips table, a drivers table, and a vehicles table. The trips table is of particular interest for this study given the inclusion of fields describing pick-up/drop-off times and locations. Trips contained in the hosted trips table have been aggregated spatially to the census tract level and temporally to the nearest 15 minutes to preserve driver and passenger privacy. Fig. \ref{img:maps} illustrates the spatial distribution of hourly average trip pick-ups in the city of Chicago in 2019. An area with high trip demand density is desired for the rebalancing study, thus the area highlighted in Fig. \ref{img:maps} was chosen for study. This area is divided into a 5-by-5 grid with grid cell size of about 1 mile by 1 mile. The grid is magnified in Fig.  \ref{img:maps} to display the routing network connecting census tract centroids.

Trip data on a randomly selected day (Friday, September 13, 2019) was used to train and test the RL rebalancing policy. This dataset contains 150,493 trips, with hourly trip pick-ups over time shown in Fig. \ref{img:hourly_trip_demand}. 
The spatial distributions of trip demands over grid cells were further explored and are presented in Fig. \ref{img:sptial_dist} which clearly illustrates the spatial shift of trip demand over time.

Several pre-processing steps were performed for the trip data to remove erroneous values and produce trip request records that more plausibly reflect real-world operations.  We:

\renewcommand{\labelenumii}{\Roman{enumii}}
 \begin{enumerate}
   \item dropped trips without defined pick-up and drop-off locations and times;
   \item dropped trips with identical pick-up and drop-off locations and times;
   \item disaggregated pick-up times by assuming trips were equally likely to occur within a 15 minute period. Application of this step retains varying trip frequency at broader time scales such as more trips at night versus early morning as shown in Fig. \ref{img:hourly_trip_demand}.
\end{enumerate}

In training the RL rebalancing policy, there is an important trade-off between model fidelity and training time because the computational time needed to run the simulator scales linearly with the number requests.  In our experiments, running a simulation with all trips for a given day led to a 60-fold increase in computational time compared to running a simulation with 1\% of the total trips.  We hypothesized, however, that using distributional action and state spaces might enable us to train the RL policy using a much smaller subset of the trip data, and then transfer this policy to a simulation using the full data set.  Of course, this hypothesis is based on the assumption that request distributions are similar between smaller and larger problem instances.  To test this hypothesis, we uniformly sampled at random 1\% ($\sim$ 1,500 trips) of the data from Friday 09/13/2019 and trained the RL policy on this scenario. We then evaluated the performance of this policy on scenarios of differing scales and days of the week, as discussed in the next section.

\subsection{Results and Discussion}\label{sec:res_disc}
\subsubsection{Fleet Sizing}\label{sssec:fleet-sizing} 
\begin{figure}[t!]
  \centering
  \includegraphics[width=0.95\linewidth]{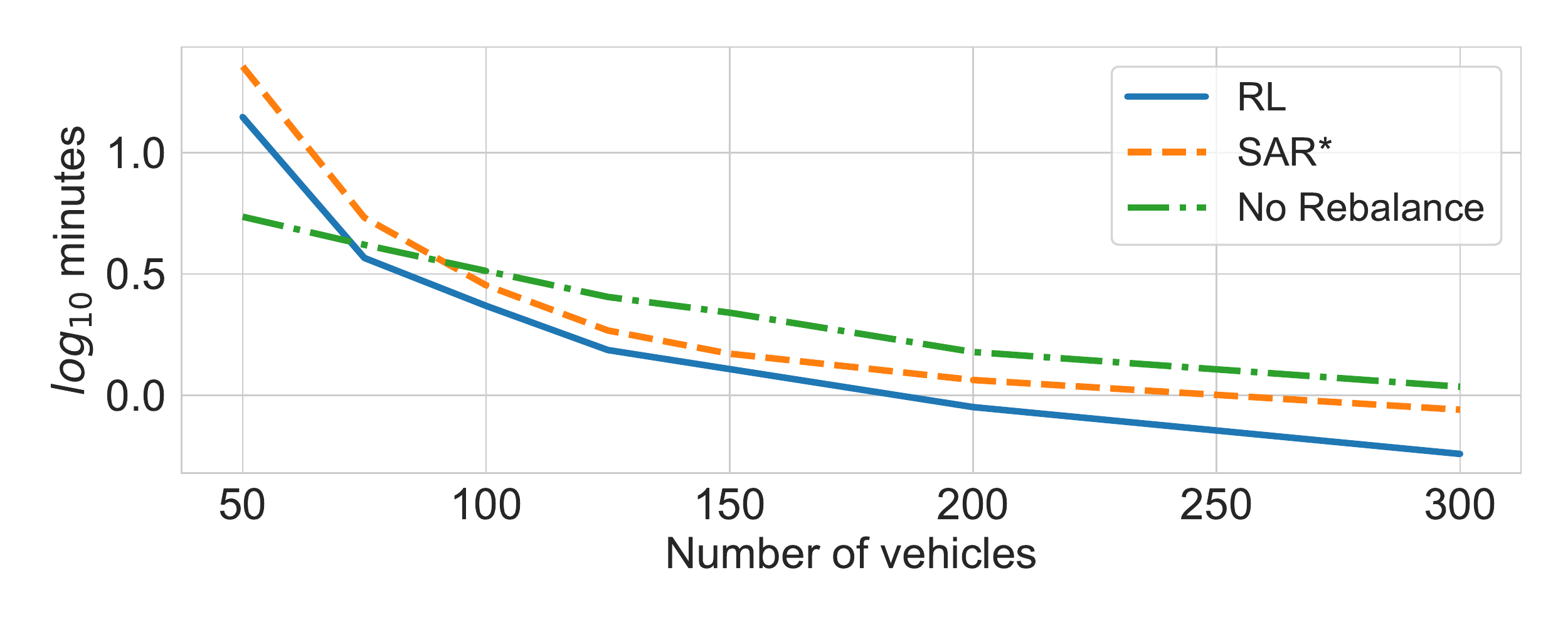}
  \caption{\footnotesize Dependence of mean wait time on fleet size for 1,500 trip requests, illustrating that rebalancing is only effective if the fleet has sufficient flexibility.  This analysis was used to select the 15-to-1 request-to-vehicle ratio used in case studies.}\label{img:sensitivity_analysis}
\end{figure}

Prior to performing our control studies, we first addressed the question of how to size the simulated vehicle fleet for the target system.  The reason for this step is that, unless the request-to-vehicle ratio is sufficiently small, the system will not have the flexibility needed to exploit a rebalancing scheme.  To establish a reasonable request-to-vehicle ratio, we performed a sensitivity analysis comparing the total wait time for 1,500 requests across different fleet sizes.  As illustrated in Fig. \ref{img:sensitivity_analysis}, both RL and \sarstar rebalancing policies are counterproductive for fleet sizes below 75 vehicles, or for request-to-vehicle ratios above 20-to-1.  Based on this observation, we used a request-to-vehicle ratio of 15-to-1 for the remainder of our case  studies.


\begin{figure}[t!]
  \includegraphics[width=\linewidth]{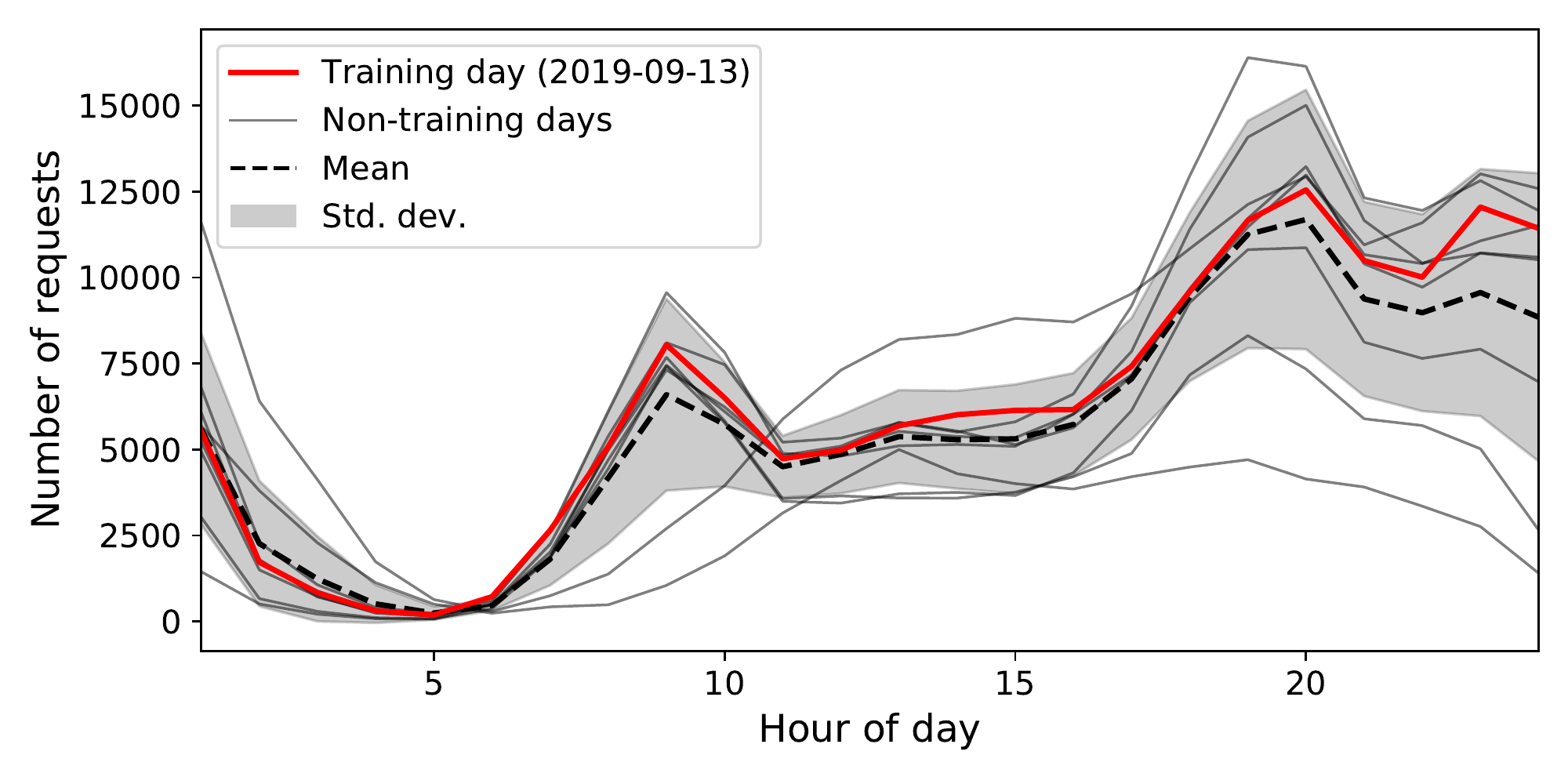}
  \caption{\footnotesize Comparison of hourly trip demand between the test dates and Friday, 09/13/2019.}\label{img:hourly_trip_demand}
\end{figure}

\subsubsection{Training day performance} 
\begin{figure}[t!]
  \centering
  \includegraphics[width=0.95\linewidth]{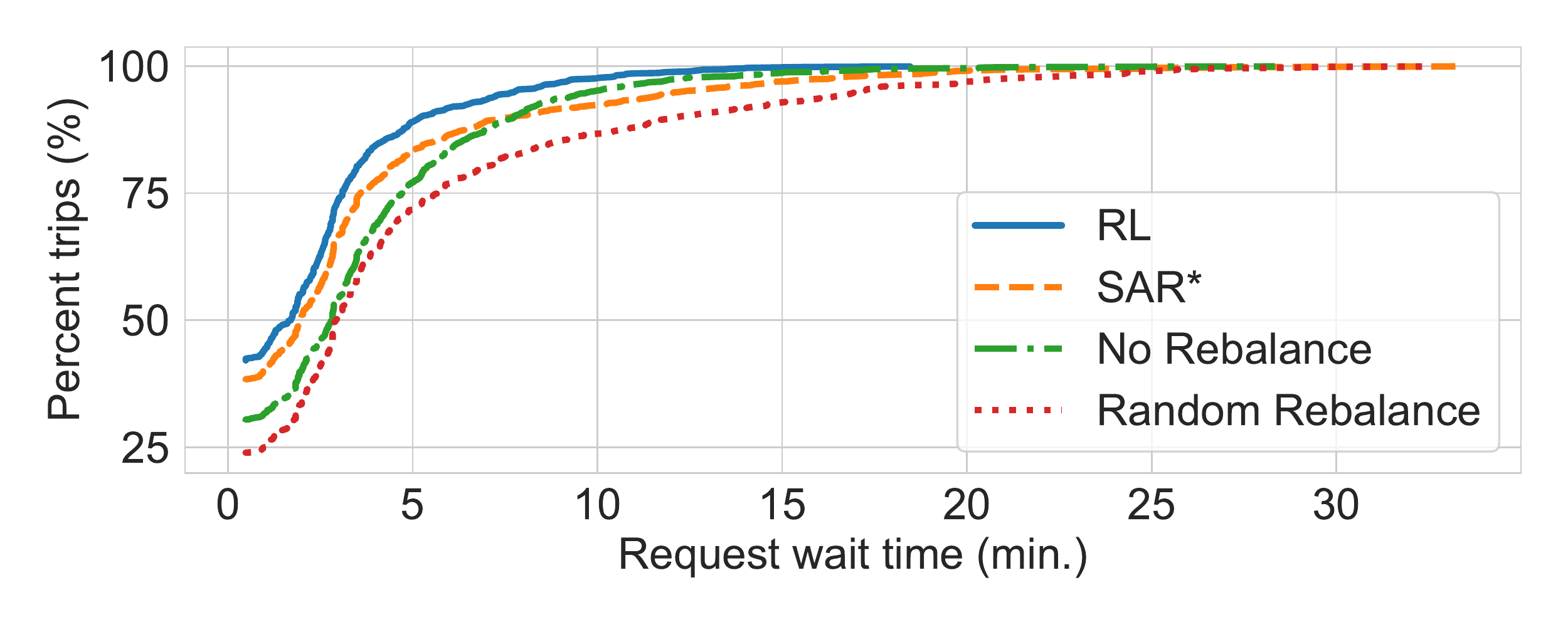}
  \caption{\footnotesize Cumulative distributions of per-request wait time for RL and the baseline algorithms on the training scenario (i.e., simulating 1,500 requests and 100 vehicles from 09/13/2019).}\label{img:total-wait-time-hist}
\end{figure}

Table \ref{tab:results1} compares the performance of the different rebalancing strategies on the training set as well as on larger-scale problem instances drawn randomly from the same date.  In addition to the baselines identified in Section \ref{ssec:baselines}, we also considered an imperfect-information version of SAR, namely  \textit{transferred SAR} (\tsar), which applies the \sarstar forecasts for 09/13/2019 with 1,500 trips and 100 vehicles as imperfect forecasts to re-scaled request predictions on the same day.  The first column of Table \ref{tab:results1} reports the mean wait time for each rebalancing algorithm, with the exception of \tsar, using a scenario of 1,500 requests and 100 vehicles.  For comparison, the difference from the no rebalancing strategy (NR) is also reported.  We see that RL, despite not having perfect information, was able to reduce the mean wait time by 28\% over NR and an additional 15\% over \sarstar.  A different view of algorithm performance is presented in Fig. \ref{img:total-wait-time-hist} which shows the cumulative distribution of request-level wait times.  Here, too, RL proves to outperform the baselines as indicated by the higher percentage of requests with the smallest wait times.  These results establish the basic efficacy of model-free RL for this problem and demonstrates that an effective (albeit approximate), simulation-driven solution to the bi-level problem (Eq. \eqref{eq-rebalancing-over-G}) is attainable. The results in Table \ref{tab:results1} and Fig. \ref{img:total-wait-time-hist} also make it clear that random rebalancing (RR) strategy is detrimental in this context, as this strategy increased the wait time by almost 50\% over no rebalancing.

\subsubsection{Same-day policy transference} We next considered the transference of the trained RL policy to problem instances with roughly 10- and 100-fold the number of requests, keeping the relative fleet size fixed at a request-to-vehicle ratio of 15-to-1 as discussed in Section \ref{sssec:fleet-sizing}.  In particular, we applied the policy learned in the 1,500-request scenario to a randomly generated scenario of 15,000 trips (sampled uniformly at random from the full data set) as well as to the full data set of 150,493 trips. Simulations were run with multiple seeds in order to provide mean and variance estimates for the reported performance.  Comparison results between RL and baselines \sarstar, \tsar, NR, and RR are presented in the second and third rows of Table \ref{tab:results1}. The results show consistent trends in waiting times across all three scenarios of 1,500, 15,000, and 150,493 trips: RL $<$ \sarstar $<$ \tsar $<$ NR $<$ RR. RL, \sarstar, and \tsar further reduced waiting times as trip number increases from 1,500 to 150,493; we speculate that this is because larger sample sizes leads to higher vehicle densities and increases the chance of same-node request-vehicle assignments. RL reduced waiting time by 28\% to 38\% compared with NR for scenarios with number of trips ranging from 1,500 to 150,493. The results indicate that the RL rebalancing policies can be indeed successfully transferred to larger problem instances with no additional training.

\subsubsection{Cross-day policy transference} The final question we address is whether the RL policy learned on one day is readily transferred to other days in the near future, as well as days from other seasons within the year. In order to objectively validate the trained policy, we consider two sets of days, namely i) other days in September 2019 (Set I) and ii) days corresponding to different seasons within the year, as well as a holiday (Set II).

\noindent\textit{Set I:} We selected three representative days to evaluate policy transfer: a weekend day (Saturday, 09/14/2019), a different weekday of the next week (Tuesday, 09/17/2019), and the same weekday of next week (Friday, 09/20/2019).  These evaluation days have 174,350 trips, 92,601 trips and 144,849 trips, respectively, with hourly trip demand patterns illustrated in Fig.  \ref{img:hourly_trip_demand}. We see in Fig. \ref{img:hourly_trip_demand} that the trip demand distribution on Saturday is qualitatively different from all other weekdays. Tuesday has similar trip demand trend to Friday but with less demand in the evening and at night. Trip demand distributions on the two Fridays (09/13/2019 and 09/20/2019) are almost indistinguishable outside of a small number of intervals.

\noindent\textit{Set II:} We also evaluated the trained policy on a set of days representing different seasons of the year, as well as on a federal holiday. Since we trained the policy on a Fall day, we selected the first three Fridays from Spring, Summer, and Winter 2019. Hence, we chose Friday 03/01/2019, Friday 06/07/2019, and Friday 12/06/2019. We also evaluated the trained policy on Labor Day 2019 (09/02/2019). The days in this set have 156,955 trips, 146,131 trips, 172,392 trips and 70,884 trips, respectively. As expected, the trip demand on Labor Day was much lower than the other days.

The RL rebalancing policy trained on 09/13/2020 was applied, without any additional training, to simulations using the full data sets on each evaluation day, and cumulative customer wait times are reported in Tables \ref{tab:results2}, \ref{tab:results3}. Wait times were also reported for \sarstar, \tsar, NR, and RR. In these simulations, \sarstar continued to use perfect forecasting for each scenario, while \tsar used the proposed actions generated by \sarstar on the training day and applied it to every new scenario. In this sense, \tsar can be thought of as using single-day historical data as an imperfect forecast for each evaluation day.

As reported in Table \ref{tab:results2}, the transferred RL policy reduced the total waiting time by 17\%, 44\%, and 34\% on Saturday, Tuesday, and Friday, respectively, compared with no rebalancing. We hypothesize that the waiting time reduction on Saturday is the smallest due to significantly different trip demand patterns compared with the training data. Tuesday has the highest waiting time reduction mainly because there are much fewer trips than Saturday and Friday. The waiting time reduction on Friday is close to that of the training day as the trip demand pattern is almost exactly the same. RL outperforms \tsar on Saturday and Friday, but yields slightly higher waiting time on Tuesday, demonstrating better temporal transferability than \tsar in general. Regarding the results for the days in Set II, as reported in Table \ref{tab:results3}, the transferred RL policy reduced the total waiting time by 23\%, 25\%, 22\%, and 5\% on March $1^{st}$, June $7^{th}$, December $6^{th}$, and Labor Day 2019, respectively, compared with no rebalancing.
For both sets, standard deviations per request were calculated to be less than 1\% of the mean values and are omitted. We note that the transferred RL policy does not perform as well as \sarstar on the evaluation days but that this is not necessarily surprising since \sarstar is still allowed to leverage perfect forecasting. The fact that transferred RL is still competitive with \sarstar is a testament to the efficacy of transfer learning within this framework. Based on these observations, it can be concluded that RL policies trained on one day using the proposed method can be adequately applied to other days in the near future as well.

\begin{table}[t!]
    \centering
    \begin{tabular}{c|cc|cc|cc}
    \hlineB{2.5}
    {} & \multicolumn{2}{c|}{\textbf{1,500/100}} & \multicolumn{2}{c|}{\textbf{15,000/1,000}} & \multicolumn{2}{c}{\textbf{150,493/10,000}} \\
    {} & Mean & $\mathcal{E}_{\text{NR}}(\%)$ & Mean & $\mathcal{E}_{\text{NR}}(\%)$ & Mean & $\mathcal{E}_{\text{NR}}(\%)$ \\
    \hlineB{2.5}
    \textbf{NR} & 3.25 & 0 & 1.79 & 0   & 1.38 & 0   \\
    \textbf{RL} & 2.34 & -28 & 1.20 & -33 & 0.86 & -38 \\
    \textbf{SAR$^*$} & 2.84 & -13 & 1.29 & -28 & 0.92 & -34\\ 
    \textbf{t-SAR} & \ditto & \ditto & 1.52 & -15 & 1.15 & -17 \\
    \textbf{RR} & 4.79 & +47  & 2.27 & +26  & 1.70 & +24 \\
    \hlineB{2.5}
    \end{tabular}
    \caption{\footnotesize Results for same-day policy transference.  Column headers indicate the request-to-vehicle ratio of scenarios sampled from the training date, Friday, September 13, 2019.  Mean customer wait time in minutes, averaged over 10 random seeds, is reported alongside percent deviation from the no rebalance strategy ($\mathcal{E}_{\text{NR}}(\%)$).}
    \label{tab:results1}
\end{table}

\section{Conclusion}\label{conclusions}
In this paper, we present a novel approach for rebalancing vehicle fleets in MoD systems. Our method differs from previous work in its ability to leverage an off-the-shelf dispatcher to effect free vehicle repositioning, as well as in the probabilistic nature of state and action spaces enabling policy transference between both problem scale and day of the week. Our method is demonstrated to be effective on a simulated system using historical request data and a data-driven vehicle routing network, providing evidence that the framework can accommodate more realistic vehicle itineraries rather point-to-point routing used in many comparable studies.  The fact that model-free RL can be successfully applied in this context suggests that it is a viable approach that does not require an all-or-nothing replacement of existing dispatch heuristics.

\begin{table}[t!]
    \centering
    \begin{tabular}{c|cc|cc|cc}
    \hlineB{2.5}
    \textbf{} &\multicolumn{2}{c|}{\textbf{Sep. 14, 2019}} & \multicolumn{2}{c|}{\textbf{Sep. 17, 2019}} & \multicolumn{2}{c}{\textbf{Sep. 20, 2019}}\\
    {} & Mean & $\mathcal{E}_{\text{NR}}(\%)$ & Mean & $\mathcal{E}_{\text{NR}}(\%)$ & Mean & $\mathcal{E}_{\text{NR}}(\%)$ \\
    \hlineB{2.5}
    \textbf{NR} & 1.13 & 0  & 0.94 & 0 & 0.87 & 0   \\
    \textbf{RL} & 0.94 & -17 & 0.53 & -44 & 0.58 & -34 \\
    \textbf{SAR$^*$} & 0.86 & -23 & 0.40& -57 & 0.46 & -47 \\
    \textbf{t-SAR} & 1.10 & -3 & 0.51 & -46 & 0.67 & -24 \\
    \textbf{RR}    & 2.04 & +81 & 1.14 & +22 & 1.36 & +55 \\ \hlineB{2.5}
    \end{tabular}
    \caption{\footnotesize Results for cross-day policy transference. The policy learned on the training date with a 1,500-to-100 vehicle-to-trip ratio was transferred to problems with request sets of sizes 174,350, 92,601, 144,849 (respectively, from left to right) and 10,000 vehicles.  Mean customer wait time in minutes, averaged over 10 random seeds, is reported alongside the percent deviation from the no rebalance strategy ($\mathcal{E}_{\text{NR}}(\%)$).}  Note that higher wait times coincide with lower vehicle-to-trip ratios.
    \label{tab:results2}
\end{table}

\begin{table}[t!]
    \centering
    \resizebox{\columnwidth}{!}{%
    \begin{tabular}{c|cc|cc|cc|cc}
    \hlineB{2.5}
    \textbf{} & \multicolumn{2}{c|}{\textbf{Mar. 1, 2019}} & \multicolumn{2}{c|}{\textbf{Jun. 7, 2019}} & \multicolumn{2}{c|}{\textbf{Dec. 6, 2019}} & \multicolumn{2}{c}{\textbf{Sep. 2, 2019}}\\
    {} & Mean & $\mathcal{E}_{\text{NR}}(\%)$ & Mean & $\mathcal{E}_{\text{NR}}(\%)$ & Mean & $\mathcal{E}_{\text{NR}}(\%)$ & Mean & $\mathcal{E}_{\text{NR}}(\%)$\\
    \hlineB{2.5}
    \textbf{NR}&1.39&0&1.25&0&1.5&0&0.79&0\\
    \textbf{RL}&1.07&-23&0.94&-25&1.18&-22&0.75&-5\\
    \textbf{SAR$^*$}&0.96&-31&0.83&-34&1.16&-23&0.62&-22\\
    \textbf{t-SAR}&1.08&-22&1.01&-19&1.3&-13&0.69&-13\\
    \textbf{RR}&2.01&+45&1.77&+42&2.34&+56&0.94&+19\\ \hlineB{2.5}
    \end{tabular}
    }
    \caption{\footnotesize Results for cross-season policy transference. The same policy used in Table \ref{tab:results2}, was also used for testing on different seasons. For objectivity, we sampled the first Fridays on March (Spring), June (Summer), and December (Winter) 2019. We also included results from testing the trained policy on a holiday (Labor Day), which was September 2, 2019.}
    \label{tab:results3}
\end{table}

In closing, we enumerate some challenges that remain in applying this framework to real-world systems, as well as proposed remedies to motivate future research.

\emph{Sample complexity and model-based RL.} Model-free RL is known to be data-intensive, which we addressed in this paper via the use of a fast, forward simulator and parallel computing for policy training.  Increasingly, model-based RL is being applied to such problems in the hope that an explicit model of the dynamics (given or learned) can act as a policy prior to reduce the sample complexity in training a new policy.  The investigation of model-based RL for fleet rebalancing constitutes an interesting research direction.

\emph{Simulation fidelity.} Our simulator ignores some important operational aspects of MoD systems such as refueling, network congestion, state estimation, and communication.  We hope to incorporate these features into future work, particularly those needed to control fleets of autonomous, electric vehicles promised to supplant conventional fleets.

\emph{Multi-objective optimization.} For illustration purposes, we used a simple cost function in this study -- mean customer wait time -- both because it encapsulates quality of service and should be directly impacted by fleet imbalance.  How to weigh this objective against other important operational metrics, such as empty vehicle miles traveled, is another important question for future research.

\emph{Policy generalization.} An interesting extension to this work would be to understand how to further generalize the transfer learning, e.g., between seasons of the year and geographic areas.  The related question of how to safely and effectively transfer an RL policy learned via simulation to a real-world twin is one that spans almost all RL research. As in this paper, the use of historical data to drive simulation should help to improve the applicability of learned policies.  It would be valuable to further investigate how, for example, a policy trained via simulation might be directly transferred to, and then tuned in relation to, the target physical system.

We hope that the present work motivates some novel ideas around integrating  model-free RL and model-based (heuristic) solution schemes to improve the state of the art in fleet rebalancing.  We believe that such hybrid methods show great promise as MoD systems evolve towards greater scale, autonomy, and electrification.



%

\appendices
\section{Proximal Policy Optimization}
PPO is very effective in the optimization of large nonlinear policies. We chose PPO due to its ability in supporting large non-linear action spaces, while guaranteeing monotonic improvement by considering the Kullback-Leibler (KL) divergence of policy updates. The main objective of PPO is to maximize the following unconstrained optimization problem:
\begin{equation}\label{eq:PPO_obj}
     \max_{\theta}\mathbb{E}_t\Bigg[\frac{\pi_{\theta}(\rlaction{t}|\rlstate{t})}{\pi_{\theta_{old}}(\rlaction{t}|\rlstate{t})}\hat{A}-\beta \text{KL}[\pi_{\theta_{old}}(\cdot|\rlstate{t}),\pi_\theta(\cdot|\rlstate{t})]\Bigg],
\end{equation}
where $\beta$ is some penalty coefficient and $\hat{A}$ is an estimator of the advantage function. PPO computes an update at each step that minimizes the cost function, ensuring that the deviation from the previous policy ($\pi_{\theta_{old}}$) is relatively small. To avoid large policy updates that could lead to training instability, PPO clips the first part of Eq. \ref{eq:PPO_obj} (also known as surrogate objective)
\[L_\theta = \mathbb{E}_t\Bigg[\frac{\pi_{\theta}(\rlaction{t}|\rlstate{t})}{\pi_{\theta_{old}}(\rlaction{t}|\rlstate{t})}\hat{A}\Bigg] = \mathbb{E}_t\Bigg[r_t^\theta\hat{A}\Bigg],\]
as:
\[L_\theta^{CLIP} = \mathbb{E}_t\Bigg[\min(r_t^\theta\hat{A}, \text{clip}(r_t^\theta, 1-\epsilon, 1+\epsilon)\hat{A}\Bigg],\]
where $\epsilon$ is a manually tuned hyperparameter, and 
\[r_t^\theta = \frac{\pi_{\theta}(\rlaction{t}|\rlstate{t})}{\pi_{\theta_{old}}(\rlaction{t}|\rlstate{t})}.\]
Algorithm \ref{alg:PPO} gives a high-level demonstration of the PPO algorithm.
\begin{algorithm}[h!]
\footnotesize
\caption{PPO algorithm}
 \begin{algorithmic}[]\label{alg:PPO}
   \FOR{iteration=1,2,...}
     \FOR{actor=1,2,...,N}
     \STATE{Run policy $\pi_{\theta_{old}}$ in environment for $T$ timesteps} 
     \STATE{Compute advantage estimates $\hat{A}_1,\hat{A}_2,...,\hat{A}_T$}
     \ENDFOR\\
     \STATE{Optimize surrogate $L^{CLIP}$ with respect to $\theta$, with $K$ epochs and minibatch size $M\leq NT$.}
     \STATE{$\theta_{old}\leftarrow \theta$}
   \ENDFOR
  \end{algorithmic} 
\end{algorithm}
For calculating the advantage function estimators $\hat{A}_{\cdot}$ we used the Generalized Advantage Estimation (GAE). In our experiments we utilized the default RLlib PPO hyperparameter configuration \cite{rllib}. The hyperpamareters and their values are summarized in Table \ref{tab:ppo_hyperparameters}.
\begin{table}[!htb]
    \centering
    \begin{tabular}{c|c|c|c}
        \hlineB{2.5}
         & Hyperparameter & Symbol & Value \\
         \hlineB{2.5}
         & Clipping range & $\epsilon$ & 0.3 \\
         & GAE parameter & $\lambda$ & 1.0 \\
         & Discount & $\gamma$ & 0.99 \\
         & Minibatch size & $M$ & 128 \\
         & Actors & $N$ & 36 \\
         & Horizon & $T$ & 4,000 \\
         & Epochs & $K$ & 30 \\
         \hlineB{2.5}
    \end{tabular}
    \caption{\footnotesize PPO hyperparameters, along with their symbols and values used for policy training.}
    \label{tab:ppo_hyperparameters}
\end{table}
The actor-critic networks have the same structure, which include two pairs of convolution and maxpooling layers, followed by a fully-connected layer that outputs the value and the proposed action. Table \ref{tab:ac_hyperparameters} gives the sizes of every network layer features.
\begin{table}[!htb]
    \centering
    \begin{tabular}{c|c|c|c}
        \hlineB{2.5}
         & Layer & Feature & Size \\
         \hlineB{2.5}
         & $1^{st}$ convolutional layer & [filters, kernel, stride] & [2, (2x2), 1] \\
         & $2^{nd}$ convolutional layer & [filters, kernel, stride] & [4, (2x2), 1] \\
         & $1^{st}, 2^{nd}$ maxpooling layer & pool & (2x2) \\
         & Fully-connected layer & units & 128\\
         \hlineB{2.5}
    \end{tabular}
    \caption{\footnotesize Neural network hyperparameters.}
    \label{tab:ac_hyperparameters}
\end{table}

\section{Simulation parameters}
Table \ref{tab:sim_hyperparameters} gives of the parameter values of our agent-based simulator as described in Section \ref{sec-simulator}. They include the various time intervals used in our simulations, as well as map grid size and the maximum passenger capacity of the vehicles, which is assumed to be the same for all of them.
\begin{table}[!htb]
    \centering
    \begin{tabular}{c|c|c|c}
        \hlineB{2.5}
         & Hyperparameter & Symbol & Value \\
         \hlineB{2.5}
         & Time step (minutes) & $\Delta t_s$ & 1.0 \\
         & Dispatch step (minutes) & $\Delta t_d$ & 1.0 \\
         & Rebalance step (minutes) & $\Delta t_r$ & 60.0 \\
         & Time horizon (minutes) & $N_s$ & 1440.0 \\
         & Maximum passenger waiting time (minutes) & $t_{wait}$ & 30.0 \\
         & Grid size & $N_x \times N_y$ & $5 \times 5$\\
         & Maximum vehicle capacity & $n_{cap}$ & 4 \\
         \hlineB{2.5}
    \end{tabular}
    \caption{\footnotesize Simulation parameters.}
    \label{tab:sim_hyperparameters}
\end{table}

\ifCLASSOPTIONcaptionsoff
  \newpage
\fi



\bibliographystyle{IEEEtran}
\bibliography{IEEEabrv,references}
%

%
\small
\begin{IEEEbiography}[{\includegraphics[width=1in,height=1in,clip,keepaspectratio]{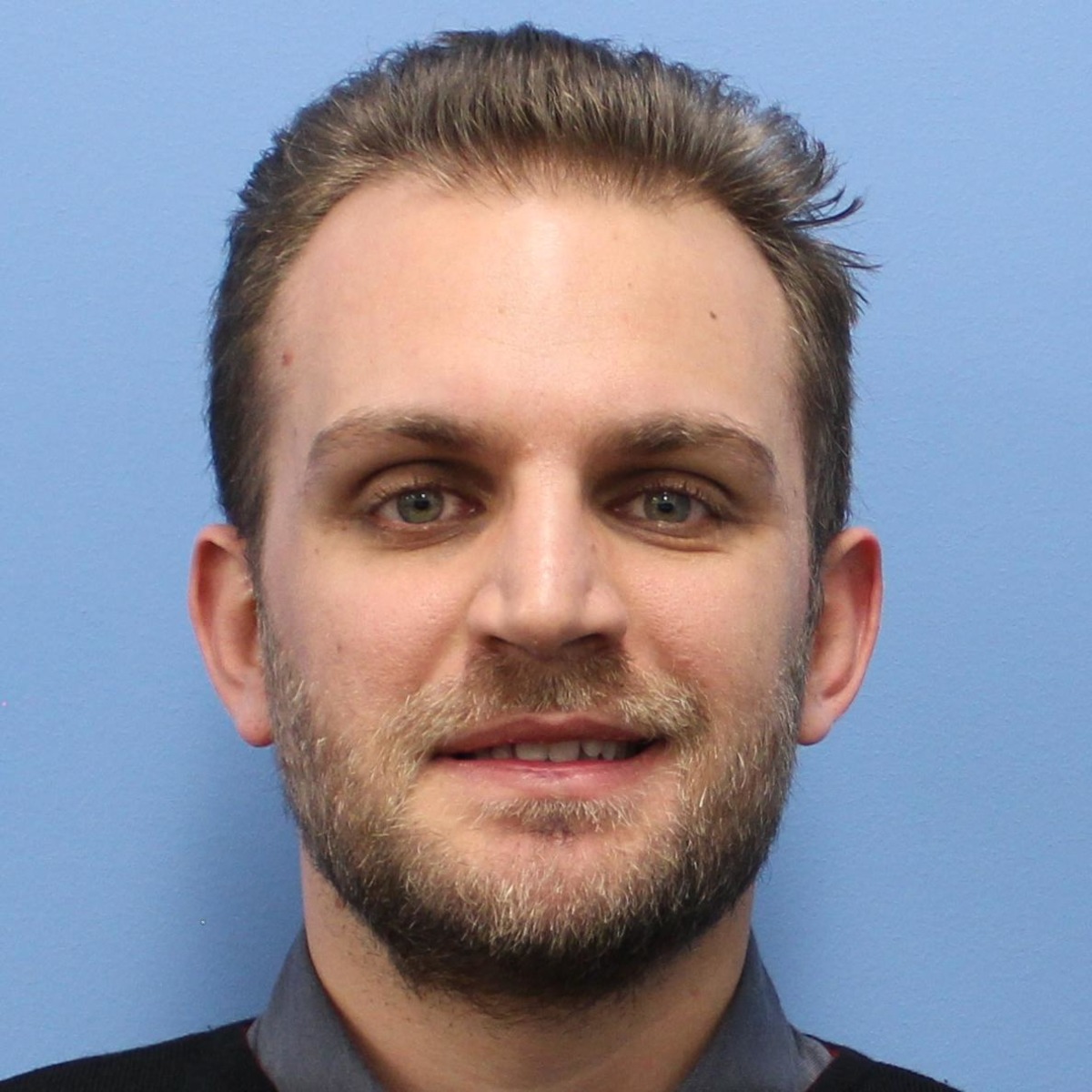}}]{Erotokritos Skordilis} is a postdoctoral researcher in the Computational Science Center at the National Renewable Energy Laboratory. He received his B.S. in computer engineering and his M.S. in mechanical engineering from University of Thessaly, Greece in 2011 and 2013, respectively, and his Ph.D. from the University of Miami in 2019. His research focuses on reinforcement learning applications in the areas of transportation and de novo molecular design.
\end{IEEEbiography}
\vskip -2\baselineskip plus -1fil
\begin{IEEEbiography}[{\includegraphics[width=1in,height=1in,clip,keepaspectratio]{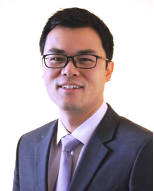}}]{Yi Hou} is a mobility science researcher at the National Renewable Energy Laboratory’s Center for Integrated Mobility Sciences (CIMS). He received M.S. and Ph.D. degrees in civil engineering from the University of Missouri in 2011 and 2014, respectively. The primary focus of his research revolves around machine learning, artificial intelligence (AI), and big data techniques, as well as their applications to future mobility and smart city solutions.
\end{IEEEbiography}
\vskip -2\baselineskip plus -1fil
\begin{IEEEbiography}[{\includegraphics[width=1in,height=1in,clip,keepaspectratio]{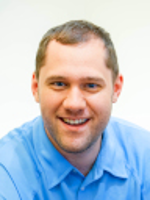}}]{Charles Tripp} received his B.S. in electrical engineering from Rice University in 2006, and his M.S. and Ph.D. degrees in electrical engineering from Stanford University in 2008 and 2013, respectively. After founding and managing a data-science startup in 2013, he drove the company to a successful acquisition and returned to research in 2019. Charles’s research focuses on advancing the frontiers of machine and reinforcement learning and its application to practical control problems.
\end{IEEEbiography}
\vskip -2\baselineskip plus -1fil
\begin{IEEEbiography}[{\includegraphics[width=1in,height=1in,clip,keepaspectratio]{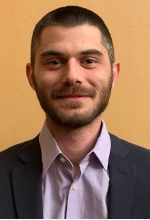}}]{Matthew Moniot} is a data analyst in the National Renewable Energy Laboratory’s Mobility, Behavior, and Advanced Powertrains (MBAP) group. He received his M.S. and B.S. degrees from Virginia Tech in Mechanical Engineering. His research primarily relates to vehicle powertrain modeling, infrastructure planning, and emerging mobility services.
\end{IEEEbiography}
\vskip -2\baselineskip plus -1fil
\begin{IEEEbiography}[{\includegraphics[width=1in,height=1in,clip,keepaspectratio]{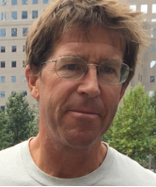}}]{Peter Graf} is a senior scientist in the Computational Science Center at the National Renewable Energy Laboratory. He received a B.S. in symbolic systems from Stanford University in 1989 and a Ph.D. in mathematics from the University of California at Berkeley in 2003. Dr. Graf’s research brings state-of-the-art applied math, computing, and AI to bear on a wide variety of renewable energy applications. His current work focuses on reinforcement learning and quantum computing for energy systems optimization and control.
\end{IEEEbiography}
\vskip -2\baselineskip plus -1fil
\begin{IEEEbiography}[{\includegraphics[width=1in,height=1in,clip,keepaspectratio]{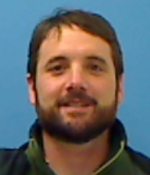}}]{David Biagioni} is a research scientist in the Computational Science Center at the National Renewable Energy Laboratory. He received his B.S. in physics from the Georgia Institute of Technology in 2004 and his Ph.D. in applied mathematics from the University of Colorado, Boulder, in 2012. Dr. Biagioni’s research combines optimization modeling, machine learning, and artificial intelligence in application to a wide range of computational problems in clean energy.
\end{IEEEbiography}







\end{document}